\documentclass[preprint]{elsarticle}

\usepackage{lineno,hyperref}
\modulolinenumbers[5]
\usepackage{subcaption}
\usepackage{algorithm}
\usepackage{algpseudocode}
\usepackage{adjustbox}

\usepackage{longtable}
\usepackage{lscape}

\usepackage{geometry}
\usepackage{pgfplots}
\usepackage{pgf-pie}  

\usepackage{setspace}
\usepackage{tikz}
\usepackage{pgfplotstable}
\usepackage{amsmath}
\usepackage{color,soul}

\date{April 2, 2022}
\journal{ }

\begin{document}
\setstretch{1.2}
\begin{frontmatter}

\title{The Use of NLP-Based Text Representation Techniques to Support Requirement Engineering Tasks: A Systematic Mapping Review}


\author[mymainaddress]{Riad Sonbol \corref{mycorrespondingauthor}}
\ead{riad.sonbol@hiast.edu.sy}

\author[mymainaddress]{Ghaida Rebdawi}
\author[mysecondaryaddress]{Nada Ghneim}

\cortext[mycorrespondingauthor]{Corresponding author}

\address[mymainaddress]{Department of Informatics, Higher Institute for Applied Sciences and Technology (HIAST)}
\address[mysecondaryaddress]{Faculty of Information and Communication Technology, Arab International University (AIU)}

\begin{abstract}
Natural Language Processing (NLP) is widely used to support the automation of different Requirements Engineering (RE) tasks. Most of the proposed approaches start with various NLP steps that analyze requirements statements, extract their linguistic information, and convert them to easy-to-process representations, such as lists of features or embedding-based vector representations.
These NLP-based representations are usually used at a later stage as inputs for machine learning techniques or rule-based methods. Thus, requirements representations play a major role in determining the accuracy of different approaches.
In this paper, we conducted a survey in the form of a  systematic literature mapping (classification) to find out (1) what are the representations used in RE tasks literature, (2) what is the main focus of these works, (3) what are the main research directions in this domain, and (4) what are the gaps and potential future directions.
After compiling an initial pool of 2,227 papers, and applying a set of inclusion/exclusion criteria, we obtained a final pool containing 104 relevant papers.
Our survey shows that the research direction has changed from the use of lexical and syntactic features to the use of advanced embedding techniques, especially in the last two years. Using advanced embedding representations has proved its effectiveness in most RE tasks (such as requirement analysis, extracting requirements from reviews and forums, and semantic-level quality tasks). However, representations that are based on lexical and syntactic features are still more appropriate for other RE tasks (such as modeling and syntax-level quality tasks) since they provide the required information for the rules and regular expressions used when handling these tasks.
In addition, we identify four gaps in the existing literature, why they matter, and how future research can begin to address them.
%The data extracted from the selected 104 papers and their categorization, can help as a useful set of references for further analysis in each task or category of tasks in RE. Besides, trends and gaps identified from this mapping study have provided many new ideas for research opportunities.

\end{abstract}


\end{frontmatter}

\section{Introduction}
\label{sec:introduction}
Requirements  Engineering (RE) is the most critical phase of the software development life cycle \cite{tahir2010requirement}. It aims to specify precisely the requirements that must be met or possessed by the desired system \cite{ieee1990ieee}. RE involves a wide range of tasks related to extracting, documenting, analyzing, validating, and managing requirements \cite{ambreen2018empirical}. The requirement represents the core elements of all these tasks. It is a broad concept describing a purpose, a need, a goal, a functionality, a constraint, a quality, a behavior, a service, a condition, or a capability \cite{jureta2008revisiting}. Requirements play a major role in the success or failure of software projects \cite{aurum2005requirements}. According to many studies, most errors in software development projects stem from the RE phase. The lack of understanding of requirements increases the risk of time and cost overrun of the project \cite{aurum2005requirements,ambreen2018empirical}.

Requirements come from various stakeholders who have different needs, roles, and responsibilities, and are as such prone to the occurrence of conflicts including interference, interdependency, and inconsistency \cite{van1998managing}. Moreover, requirements typically are specified in natural languages, which increases the complexity of requirement engineering because of the inherent ambiguity, incompleteness, and inaccuracy of natural languages \cite{denger2003higher}. These factors make RE tasks challenging, time-consuming, and error-prone work mainly for large projects, as large volumes of requirements need to be processed, analyzed, and understood \cite{vlas2011rule}.

Many researches have been carried out on the automation of different RE tasks. The proposed approaches usually start by applying a set of Natural Language Processing (NLP) steps that extract linguistic features and information from requirements texts and construct various NLP-based representations. These representations are used in further stages to solve the targeted task (e.g., classifying requirements \cite{kurtanovic2017automatically, dalpiaz2019requirements, hey2020norbert}, detecting trace links, discovering quality defects, etc.).

In the last decade, a growing number of works employed various syntactic and semantic-based features to represent requirements. The used representation is considered one of the most important factors that affect the accuracy of the proposed solutions for different RE tasks  \cite{ferrari2017natural, garigliano2018semantics}. Therefore, there is a need to analyze the related literature to investigate the suggested representations, identify gaps, and guide future research.

To address the aforementioned needs, a systematic mapping review was used as a research method for this study. A mapping review is aimed at providing an understanding of the scope of the research activities in a given area \cite{booth2021systematic}. Compared to a traditional literature review, a mapping review has many advantages, such as a well-defined methodology that reduces bias and a wider context that allows more general conclusions \cite{petersen2015guidelines}.
The review presented in this paper aims to identify all recently published primary literature that employs NLP-based representations to support RE tasks. We classify these works based on two aspects: the targeted RE task and the proposed representation, and identify potentials and gaps in the field to inform future research.

The rest of the paper is structured as follows: In section 2 we provide a background for NLP-related concepts. Section 3 provides a quick overview of related works. Section 4 describes the methodology adopted to conduct this study including the search terms, online databases, and the systematic mapping process. Section 5 presents and discusses our results. We summarize our findings in section 6 and conclude our paper in section 7.

\section{Background: Natural Language Processing}
Natural language processing is one of the main artificial intelligence disciplines. It aims to enable computer programs to “understand” and process natural language texts to achieve some specific goals \cite{jurafsky2000speech, teller2000speech}. Traditionally, there are three main levels of processing in an NLP-based approach: \cite{jurafskyspeech}: lexical and morphological level, syntactic level, and semantic level.

\subsection{Morphological (or Lexical) Level}
 The Morphological level focuses on analyzing words into their morphemes like prefixes, suffixes, and base words. It includes common tasks, such as Tokenization and Lemmatization \cite{jurafskyspeech}.

 \vspace{5pt}
 \textbf{Tokenization:} the process of splitting a text into a list of tokens. Tokens can be words, numbers, or punctuation marks.
 
 \vspace{5pt}
 \textbf{Lemmatization:} the process of finding the dictionary form, or the lemma, of each word. For example, the lemma of \emph{“Supporting”} and \emph{“Supported”} is \emph{“Support”}. 
 
\subsection{Syntactic Level} 
 The Syntactic level focuses on analyzing the grammatical structure of sentences. This level usually includes Part-of-Speech Tagging (POS-tagging), Chunking, dependency Parsing, and Named-Entity Recognition  \cite{jurafskyspeech}.
 
 \vspace{5pt}
 \textbf{POS-Tagging:} the process of tagging each token in a sentence with its corresponding part of a speech tag (such as noun, verb, adjective, etc.) based on its syntactical context   \cite{toutanova2003feature}.
 
 \vspace{5pt}
 \textbf{Chunking:} the process of detecting syntactic constituents such as Noun Phrases and Verb Phrases in a sentence   \cite{abney1996partial}. 
 
 \vspace{5pt}
 \textbf{Dependency Parsing:} the process of analyzing the syntactic structure of the sentence, by finding out the grammatically related words, as well as the type of the relationship between them   \cite{de2008stanford}.
 
 \vspace{5pt}
 \textbf{Named-Entity Recognition:} It seeks to locate and classify named entities mentioned in the sentence into pre-defined categories such as person names, organizations, locations, etc \cite{nadeau2007survey}.

\subsection{Semantic Level}
The Semantic level focuses on understanding the meaning of the text. The main goal of semantic processing is to automatically map a natural language sentence into a formal representation of its meaning. Different semantic representations have been proposed in the literature, such as:

\vspace{5pt}
 \textbf{Ontology-Based Representation:}
The ontology is a data model that represents a set of concepts within a domain and the relationships between those concepts \cite{gruber1993translation}. WordNet \cite{miller1995wordnet} is one of the widely used lexical ontologies. Ontologies are commonly used to assign words to a predefined set of concepts and to measure semantic similarity between them using different ontology-based measures such as Wu and Palmer \cite{wu1994verb} and path-based similarity \cite{rada1989development}.

\vspace{5pt}
\textbf{Vector Space Model (VSM):} A basic representation model that represents text as a term-by-document matrix \cite{salton1988automatic}. The Bag-of-Words (BOW) model is a special case for VSM where words frequencies are used as weights, and words are used as features. Other weighting factors are used in VSM, such as IDF and TF-IDF. 

\vspace{5pt}
\textbf{Topic Modeling-Based Representation:} A statistical modeling approach used to discover the latent or abstract topics that occur in a set of texts. These topics are used to represent each text. This approach helps in finding a low-rank approximation to the term-document matrix by retaining the semantic relations between words. Latent Dirichlet Allocation (LDA) \cite{blei2003latent} and Latent Semantic Analysis (LSA) \cite{dumais2004latent} are two common examples of this approach.

 \vspace{5pt}
%Prediction Based Word Embedding Techniques
\textbf{Advanced Embedding Techniques:} Embedding is an efficient method for learning high-quality vector representations of words from large amounts of unstructured text data \cite{mikolov2013efficient}. Word embedding can capture the context of a word within a document, which allows words with similar meanings to have similar vector representations.
Many famous pre-trained word embeddings are available to the public, such as Word2Vec \cite{mikolov2013efficient}, GloVe \cite{pennington2014glove}, BERT \cite{devlin-etal-2019-bert}, etc. 
 
\section{Related Reviews}
Many reviews have been published on the relation between NLP and RE tasks \cite{zhao2021natural,raharjana2021user,amna2022ambiguity}. 

Some of these reviews  provide a broad picture of NLP activities and technologies used in the RE domain \cite{dermeval2014systematic,nazir2017applications,zhao2021natural}.
Dermeval et al. \cite{dermeval2014systematic} conducted a systematic literature review to identify the primary studies on the use of ontologies in the RE domain. This study considered 77 studies published between January 2007 and October 2013.
Nazir et al. \cite{nazir2017applications} investigated the applications of NLP in the context of RE. It included 27 studies published between 2002 and 2016.
Zhao et al. \cite{zhao2021natural} introduced a comprehensive overview of the applications of NLP in RE research focusing on the state of the literature, the state of empirical research, the research focus, the state of the practice, and the NLP technologies used. This study reviewed 404 relevant primary studies reported between 1983 and April 2019.

\vspace{5pt}
 
Other reviews focused on specific RE problems. Bozyigit et al. \cite{bozyiugit2021linking} provided a review of 44 primary studies related to the automatic transformation of software requirements into conceptual models. It covered works published between 1996 and 2020.

\vspace{5pt}
 
On the other hand, a number of reviews have limited their work to specific templates for requirements \cite{amna2022ambiguity,raharjana2021user}.
Amna et al. \cite{amna2022ambiguity} reviewed studies that investigate or develop solutions for problems related to ambiguity in user stories. The study covered 36 researches published between 2001 and 2020. Similarly, Raharjana et al. \cite{raharjana2021user} presented a systematic literature review for research related to the role of NLP on user story specification. This work found 30 primary studies between January 2009 to December 2020.

Although all these works provided good information regarding requirements engineering, no conducted secondary studies have focused on the used techniques to represent textual requirements and the involved syntactic and semantic aspects in these representations.

\section{Research Method}
This study was undertaken as a systematic mapping review using the guidelines presented in Petersen et al. \cite{petersen2015guidelines}. The goal of this review is to identify, categorize, and analyze existing literature published between 2010 and 2021 and use syntactic and semantics aspects to represent software requirements when handling RE tasks.

\subsection{Planning}

\begin{table*}
\begin{tabular}{p{0.07\linewidth} p{0.20\linewidth} p{0.65\linewidth}}
\hline
Group & Main Keyword & Enriched List\\
\hline
Group A & Requirements Engineering & "requirement engineering" AND "software requirement"\\
\\
Group B & Syntactic Processing & "syntax" OR "POS-tagging" OR "tagging" OR "Dependency Parsing" OR "shallow parsing" OR "chunking" OR "named entity recognition"\\[6pt]
\\
Group C & Semantic Processing & "semantic" OR "BERT" OR "word embedding" OR "word2vec" OR "Vector Space Model" OR "VSM" OR "Latent Semantic Analysis" OR "LSA"\\
\hline
\end{tabular}
\caption{Keywords used in our study} \label{tkeywords}

\end{table*}

In this section, we define our research questions, the search strategy we use, and the inclusion and exclusion criteria considered to filter the results.

\subsubsection{Research question}

Our work is guided by the following main research questions: 

\textit{“How are the syntactic and semantics aspects considered to represent software requirements when handling RE tasks?”.}

This question is detailed in the following five RQs:
\begin{enumerate}[(RQ1)]
\item What is the state of the published literature on RE works that use syntactic and semantic representations for requirements?
 \item In which RE tasks are the syntactic and semantic aspects mostly considered to represent requirements?
 \item What are the proposed representations in the literature for RE tasks?
 \item What are the main research directions to represent requirements for each category of RE tasks?
 \item What gaps and potential future directions exist in this field?
\end{enumerate}

While the first question (RQ1) focuses on having a general overview of the published works (number of publications and the top publication venues), the second and the third questions analyze the targeted problems in RE (RQ2) and the proposed solutions (RQ3). The fourth research question (RQ4) explores the current research directions for each category of RE tasks, and the fifth question (RQ5) discusses the gaps and possible future improvements in this domain.

\subsubsection{Search Strategies}

We used five digital libraries to conduct the automated search: Scopus, IEEE Xplore, ACM Digital Library, ScienceDirect, and SpringerLink. Scopus and ScienceDirect are general indexing systems that help to cover a broader scope for our search. On the other hand, IEEE Xplore, ACM Digital Library, and SpringerLink publish papers related to the most well-known conferences and journals related to the software domain. 
 
The search string is built based on the following three key terms: “\textit{Requirements Engineering}”, “\textit{Syntactic Processing}”, and “\textit{Semantic Processing}”. These terms are derived from our main research question. Each of these terms is enriched by adding synonyms and sub-fields. Table \ref{tkeywords} shows the whole set of selected keywords for this study divided into three groups: A, B, and C.
These groups were used to create the final search query as follows:
\begin{center}
 \textit{ A \textbf{AND} ( B \textbf{OR} C) }
\end{center}

\subsubsection{Inclusion and Exclusion Criteria}
Inclusion and exclusion criteria are used to filter out papers that are not relevant to our research questions. We defined three inclusion criteria and four exclusion criteria.

\vspace{5pt}
 
\textbf{Inclusion Criteria:} 
\begin{enumerate}[IC1:]
 \item Peer-reviewed research presents an approach related to the field of software requirements engineering.
 \item The research uses NLP techniques including syntactic or semantic processing. 
 \item It is published between 2010 and 2021 
\end{enumerate}

\vspace{5pt}
 
\textbf{Exclusion Criteria:}
\begin{enumerate}[EC1:]
 \item Secondary research is excluded (such as literature reviews, summaries, etc.)
 \item The research is published in languages other than English
 \item Duplicate papers (only the most recent and detailed one is considered)
 \item The study does not provide detailed information about the proposed approach (such as, short papers, posters, etc.)

\end{enumerate}

\subsection{Conducting the Search}

\begin{figure}
 \centering
 \includegraphics[width=0.4\textwidth]{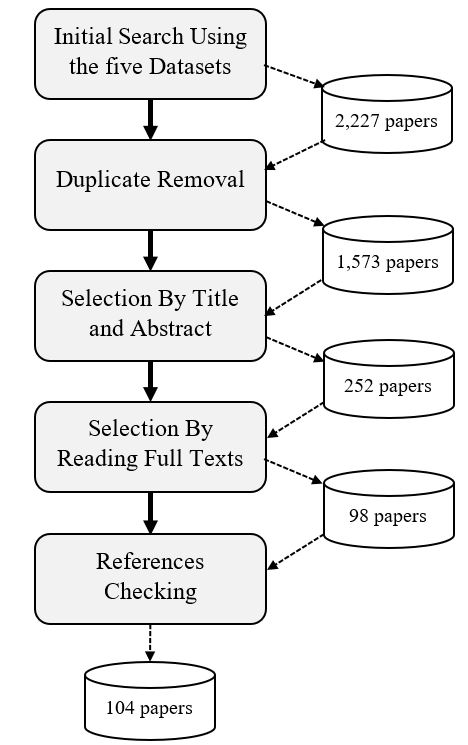}
 \caption{Search flow diagram for our systematic review}
 \label{fig:search_flow}
\end{figure}

The review process consists of five main stages. Figure \ref{fig:search_flow} illustrates these stages and the numbers of selected publications after each stage. Starting from the defined data sources, we obtained a total of 2,227 candidate papers. Duplicated papers were automatically eliminated using
Parsifal tool\footnote{https://parsif.al} first, and then additional duplicate entries were manually eliminated by comparing authors, titles, and abstracts.
After removing all duplicates, 1,573 papers remained.

These papers are filtered by applying the selection criteria based on titles and abstracts. This stage led to  the selection of 252 papers. A full-text review of those papers was then conducted to discard papers that did not satisfy our selection criteria. The remaining primary research papers after this stage are 98. Finally, reference checking led to an additional 6 relevant papers. After these stages, our final list consists of 104 unique primary papers.

\subsection{Data Extraction and classification}

A Data Extraction Form (DEF) is developed to collect the required data to answer our RQs. The form is designed in a table format consisting of the following types of information:
\begin{itemize}
 \item Bibliometric Information: author(s), publication year, type of publication, and publication venue.
 \item Targeted RE task(s).
 \item Proposed solution including the Syntactic and semantic information used to represent requirements.
 \item Evaluation Details: evaluation dataset, used metrics, and the results.
 \item limitations and constraints.
\end{itemize}

\section{Results and Discussions}
This section describes and explains the results obtained by the analysis of the 104 selected papers answering the RQs exposed in the previous sections.

\subsection{(RQ1) What is the state of the published literature on RE works that use syntactic and semantic representations for requirements?}
\vspace{5pt}
 
\begin{figure}
 \centering
 \includegraphics[width=0.5\textwidth]{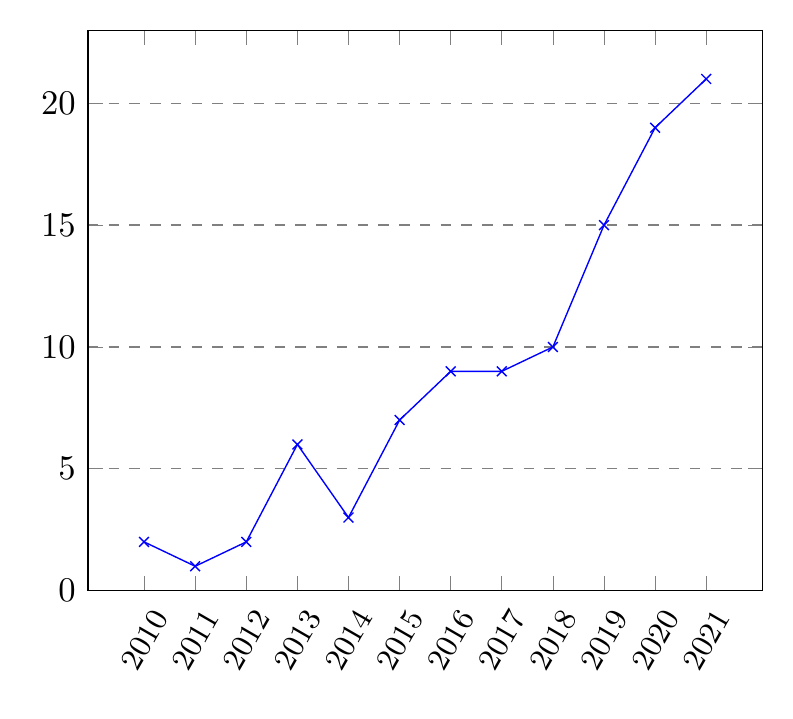}
 \caption{Number of published per year}
 \label{fig:per_year}
\end{figure}

 As previously mentioned, our final list consists of 104 unique peer-reviewed research papers. Fig \ref{fig:per_year} shows the distribution of these publications per year. This chart reflects a growing interest in the use of syntactic and semantic levels of processing to handle RE tasks. The vast majority of papers (more than 85\%) have been published since 2015.

Among the selected papers, 47\% of them appeared in journals; 42\% of papers were published in conference proceedings, while 11\% of papers came from workshops.

The most popular venue for publishing articles related to our study is \textit{IEEE International Conference on Requirements Engineering} and its workshops with more than 12\% of papers, while the most popular journals are: \textit{Information and Software Technology}, \textit{Empirical Software Engineering}, \textit{Requirements Engineering} with 5-6\% of papers in each of them.

\subsection{(RQ2) In which RE tasks are the syntactic and semantic aspects mostly considered to represent requirements?}
\vspace{5pt}

\begin{figure}
 \centering
 \includegraphics[width=1.0\textwidth,trim={0.1cm 5cm 0.5cm 3.5cm},clip]{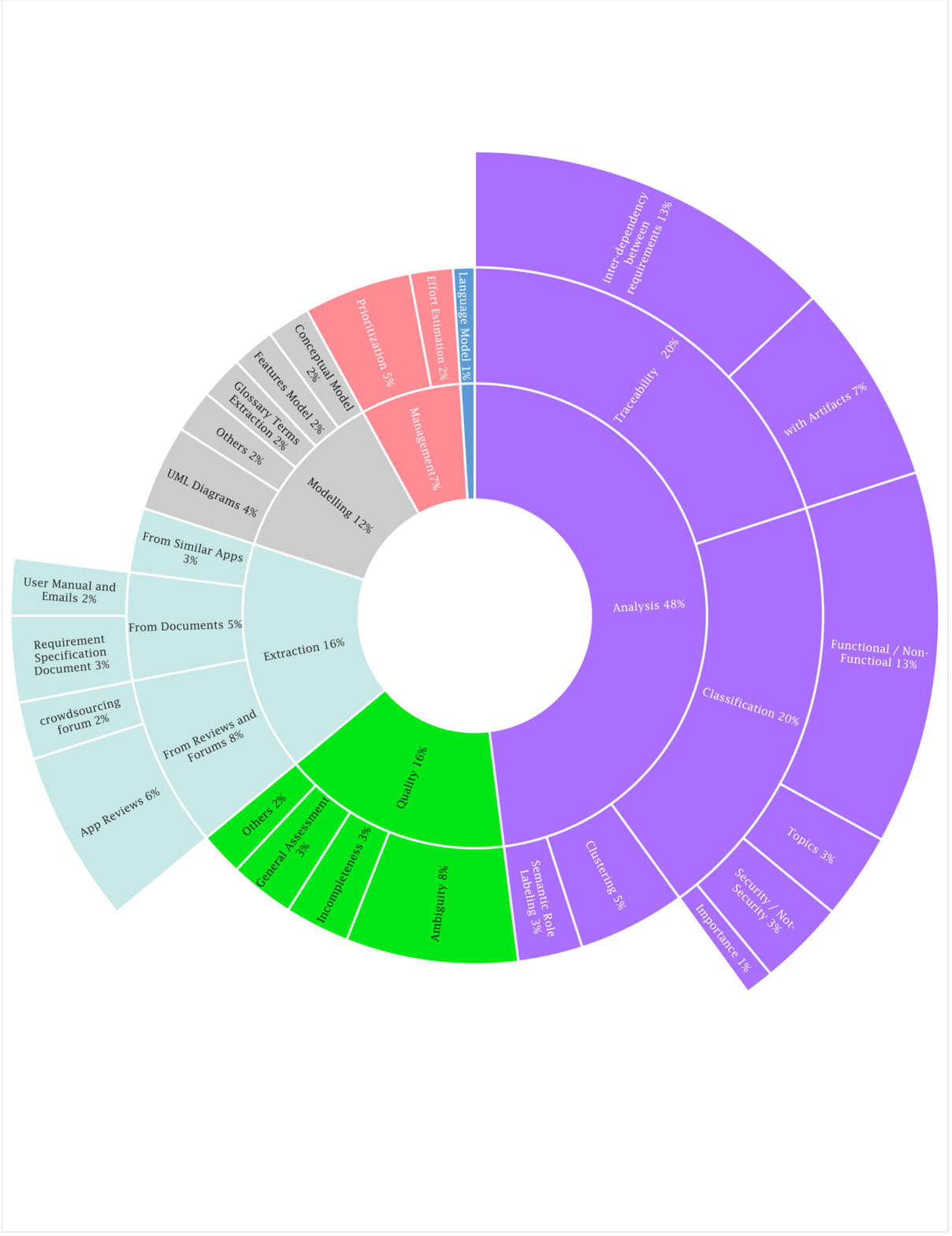}
 \caption{A chart illustrating the hierarchy of papers categorized based on their targeted RE task.}
 \label{fig:treemap}
\end{figure}

We classified the retrieved papers based on the type of the targeted RE task. Fig. \ref{fig:treemap} shows the hierarchy of all 104 papers categorized into 6 top-level categories and 19 subcategories. The six main categories are:

\subsubsection{Requirements Analysis}
This category represents the majority of publications (50 out of 104 papers). It includes papers focusing on the following RE tasks:

\begin{itemize}
    \item \textbf{Requirements Classification:}
    Requirements classification is an important step towards automatically analyzing natural language requirements, especially when handling projects with large numbers of requirements \cite{dalpiaz2019requirements}. We recognized 21 papers proposing solutions to various requirements classification tasks. Most of them focused on Functional/Non-Functional classification tasks \cite{baker2019automatic,dias2020software,casamayor2010identification,dalpiaz2019requirements,hey2020norbert,khatian2021comparative,rahman2019classifying,rashwan2013ontology,younas2020extraction,mahmoud2016detecting,kicibert,raharja2019classification,krasniqi2021analyzing}, while the remaining focused on other classification tasks: security/Not security \cite{palacio2019learning,kobilica2020automated,li2020ontology}, topic-based classification \cite{ott2013automatic}, and classification based on requirements importance level \cite{chatterjee2020identification}.
    
    \item \textbf{Requirements Traceability:}
    Requirements traceability is used to capture the relationships between the requirements, the design, and the implementation of a system \cite{wang2021analyzing}. We recognized 21 papers related to this task, part of which focused on detecting the relationships between requirements (inter-dependency between requirements) \cite{das2021sentence,blake2015shared,sonbol2020towards,alhoshan2019using,deshpande2019data,samer2019new,sultanov2011application,lu2017automatic,deshpande2021bert,aldekhail2022intelligent,nicholson2021issue}, while the remaining focused on detecting the relationships between requirements and other artifacts (design documents and source code) \cite{guo2017semantically,wang2021analyzing,wang2020automated,rasekh2019mining,mahmoud2015role,ali2019exploiting,leitao2021srxcrm,chhabra2017requirements,arora2015change}.

    \item \textbf{Requirements Clustering:} 
    Requirements clustering is used to organize software requirements into a set of clusters with high cohesion and low coupling \cite{al2005toward}. We recognized 5 papers presenting approaches to cluster requirements. These papers used the resultant clusters to understand the main functional groups or topics over requirements \cite{gulle2020topic,casamayor2012functional,misra2016topic}, to organize requirements in a tree structure (hierarchy) \cite{eyal2018semantic}, or as a step towards discovering redundancy and inconsistency between requirements \cite{mezghani2018industrial}.
    
    \item \textbf{Semantic Role Labeling:} 
    Semantic role labeling is the task of extracting semantic information from a software requirements specification \cite{wang2015semantic}. Four papers addressed this task \cite{diamantopoulos2017software,wang2016automatic,wang2015semantic}. These works focused on mapping requirements to formal representations by extracting their main semantic elements such as actors, actions, and objects.
\end{itemize}

\subsubsection{Requirements Extraction}
Requirements extraction (or elicitation) is one of the crucial steps in software development. We recognized 19 papers addressing tasks related to this category. These papers can be divided into three groups:
\begin{itemize}
    \item \textbf{Extracting Requirements from Reviews and Forums:} 9 papers \cite{de2021re,bakar2016extracting,carreno2013analysis,jha2018using,li2018automatically,lu2017automatic,peng2016approach,yang2021phrase,wu2021identifying} proposed solutions for this task, where various applications reviews and forums are used as a source for input texts such as App Store and Google Play.

    \item \textbf{Extracting Requirements from Textual Documents:} 7 papers \cite{shi2021automatically,wang2013automatic,quirchmayr2018semi,haris2020automated,abualhaija2020automated,dollmann2016and,sainani2020extracting} presented approaches to extract requirements from SRS documents, policies, user manuals, and emails.
    
     \item \textbf{Extracting Requirements from Similar applications:} 3 papers \cite{jiang2019recommending,abbas2020automated,do2020capturing} focused on recommending requirements based on the specifications of similar applications. This goal is achieved by processing the description of similar products to suggest new possible features and generate creative requirements.
\end{itemize}

\subsubsection{Quality assessment}
Quality assessment tasks are concerned with detecting defects in software requirements specifications.  \cite{femmer2017rapid,ferrari2018detecting}. We recognized 17 papers focusing on tasks related to the following quality assessment tasks:
\begin{itemize}
    \item \textbf{Ambiguity Detection:}  This task helps in the identification of ambiguous requirements. The works under this sub-category (9 papers) can be further classified based on the discussed level of ambiguity:
    \begin{itemize}
        \item Lexical Ambiguity: The main focus at this level is the ambiguity caused by words and terms. We recognized 6 papers related to this level \cite{wang2013automatic,mishra2019use,matsuoka2011ambiguity,ferrari2019nlp,dalpiaz2019requirements,misra2013entity}.
        
        \item Syntactic Ambiguity: This level focuses on detecting sentences that have different possible grammatical structures. We recognized one paper handling this level of ambiguity (Osama et al. \cite{osama2020score}).
        
        \item Semantic Ambiguity: This level focuses on detecting confusing contexts in sentences such as anaphoric ambiguity and coordination ambiguity. We recognized one paper handling this level (Ezzini et al. \cite{ezzini2021using}).
        
        \item Pragmatic Ambiguity: This level focuses on detecting sentences with multiple meanings. We recognized one paper handling this level (Ferrari et al. \cite{ferrari2012using})
    \end{itemize}
    
    \item \textbf{Incompleteness Detection:} This task is  concerned with detecting any possible incompleteness in requirements statements. We recognized 3 papers handling this task \cite{ferrari2014measuring,liu2021automated,baumer2018flexible}.
    
    \item \textbf{Conformance With Templates:} This task focuses on verifying that the requirements are indeed written according to pre-defined templates. We recognized one work proposing a solution to handle this task (Arrora et al. \cite{arora2015automated}); specifically, this work focused on checking the conformance of requirements with two well-known templates: Rupp \cite{rupp2011requirements} and EARS \cite{mavin2009easy} templates. 
    
    \item \textbf{Vagueness Detection:} 
    Vagueness occurs when a statement can have a continuum of interpretations (e.g., when using words like tall, large, etc.). We recognized one paper focusing on this problem (Cruz et al. \cite{cruz2017detecting}).
    
    \item \textbf{General Assessment:} Other papers suggest approaches providing a general assessment for different aspects of requirements quality. Three papers can be classified under this type \cite{parra2015methodology,femmer2017rapid,ferrari2018detecting}
\end{itemize}

\subsubsection{Modeling}
Modeling software requirements is the process of transferring the natural language requirements into
models and diagrams \cite{elallaoui2018automatic}. We recognized 12 papers proposing solutions related to this type of RE tasks. One of the main differences between these works is the type of the generated model:
\begin{itemize}
    \item Use Case Diagrams \cite{tiwari2019approach,hamza2019generating,elallaoui2018automatic,al2018use}, that focus on actors and their corresponding actions.
    
    \item Feature Models \cite{sree2018extracting,hamza2015recommending}, that define features and their dependencies.
    
    \item Conceptual Diagrams\cite{thakur2016identifying,lucassen2017extracting}, that focus on concepts and relationships between them.
    
    \item Glossaries \cite{arora2016automated,bhatia2020clustering}, which define technical terms  which are specific to an application domain.
    
    \item Goal Models \cite{gunecs2020automated}, that focus on the objectives which a system should achieve through the cooperation of actors in the intended software and the environment.
    
    \item Semantic of Business Vocabulary and Rules (SBVR) \cite{haj2021semantic}, which defines the semantics of business vocabulary, business facts, and business rules.
\end{itemize}

\subsubsection{Requirements Management}
Requirements management is an ongoing activity throughout the development process \cite{tzortzopoulos2006clients}. We classified two tasks under this category:
\begin{itemize}
    \item \textbf{Requirements Prioritization:} the main target for this task is to determine which candidate requirements of a software product should be included in a certain release. We recognized 5 papers handling this task \cite{ali2021requirement,kifetew2021automating,mczara2015software,misra2014latent,shafiq2021nlp4ip}.
    
    \item \textbf{Effort Estimation:}
    This task focuses on estimating the effort involved in implementing a requirement \cite{choetkiertikul2018deep}. We found 2 papers proposing methods to handle this problem \cite{hussain2013approximation,choetkiertikul2018deep}. Both of them focused on estimating the effort in Agile development methodologies.
\end{itemize}

\subsubsection{Others} 
One of the retrieved papers (Mishra et al. \cite{mishra2021crawling}) focused on building a language model for the software requirement domain. This model was based on a domain-specific text corpus collected by crawling the software engineering category on Wikipedia.

\subsection{(RQ3) What are the proposed representations in the literature for RE tasks?}

In general, almost all covered papers consist of two main phases:
\begin{enumerate}[(1)]
 \item Representation Phase: NLP processing steps are applied to analyze requirements texts, and to capture linguistic information in order to represent them in various forms.
 \item Solving Phase: the results of the previous stage are used to solve the targeted problem based on various ML and non-ML approaches.
\end{enumerate}

\begin{figure}
 \centering
 
 \includegraphics[width=0.9\textwidth]{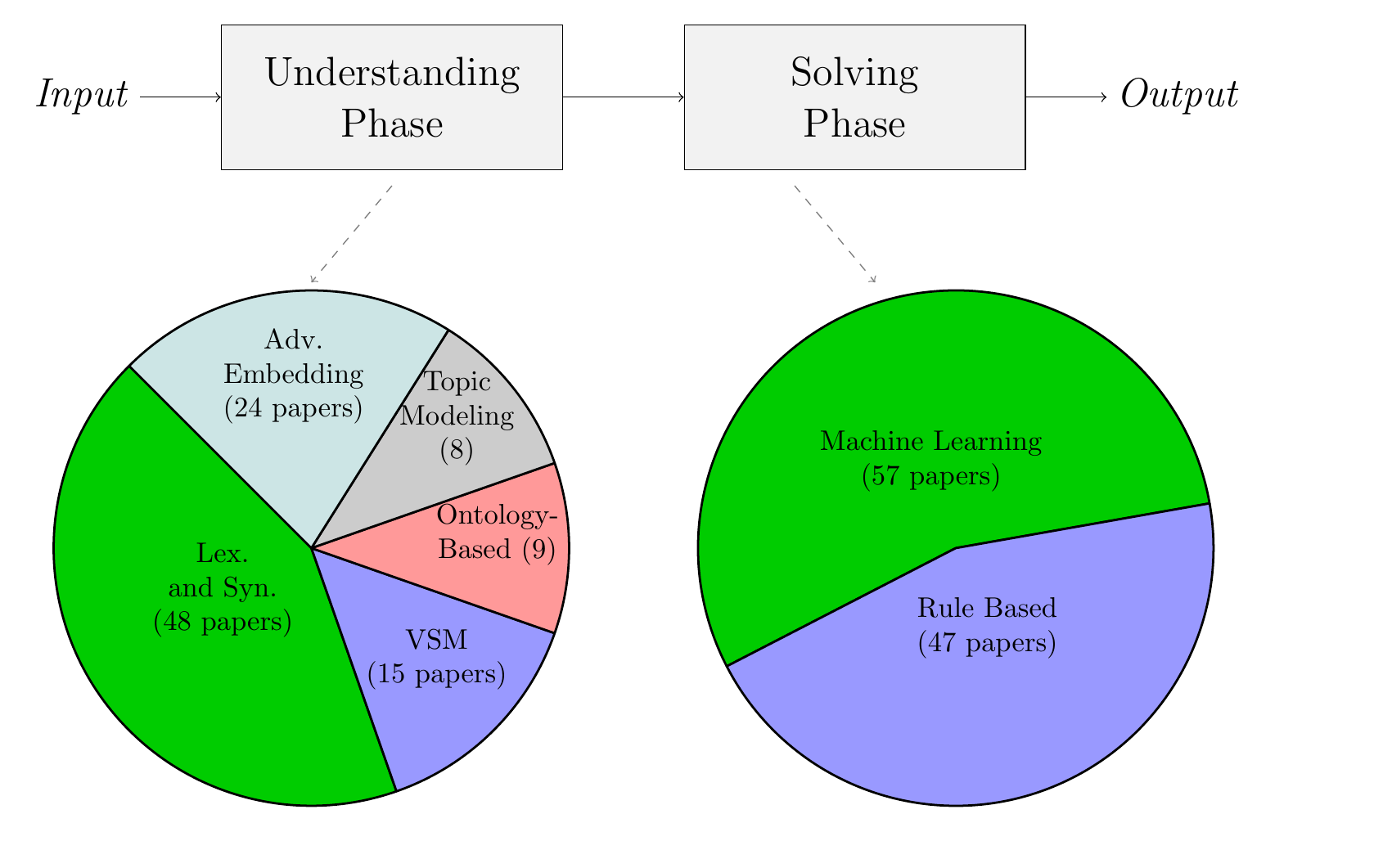}
 
 \caption{The general flow and the used techniques for each phase}
 \label{fig:general_flow}
\end{figure}

Fig. \ref{fig:general_flow} shows these two phases and the techniques used for each of them. Overall, we recognized \textbf{five} different text representation techniques used to handle the first phase: Lexical and syntactic features, ontology-based representation, VSM, topic modeling-based representation, and advanced embedding techniques. On the other hand, two families of solutions were proposed for the second phase: Machine Learning approaches which represent 55\% of papers, and Rule-Based approaches (in 45\% of papers) where patterns, regular expressions, and heuristics were used.

To answer our third research question (RQ3), we explore the representations used in the first phase in more detail:

\subsubsection{Lexical and Syntactic features}
 This part represents the largest group with more than 46\% of papers (48 out of the selected 104 papers). The solutions proposed in these papers share a similar pipeline: (1) Applying a set of NLP pre-processing steps. (2) Representing each requirement as a pre-defined set of linguistic features. (3) Then, proceeding to the solving phase which uses ML techniques (such as Decision Trees \cite{dollmann2016and,hussain2013approximation,wang2015semantic,parra2015methodology,li2020ontology}, SVM \cite{ott2013automatic,dalpiaz2019requirements,li2018automatically}, and RF \cite{abualhaija2020automated,deshpande2019data,yang2021phrase,krasniqi2021analyzing}), or rule-based approach using a set of syntactic regular expressions \cite{thakur2016identifying,haris2020automated,lucassen2017extracting,haj2021semantic,cruz2017detecting}.
 
 The used pre-defined set of features (in step 2) usually includes features related to the following four groups:

  \begin{figure}[h]
     \centering
     \includegraphics[width=0.6\textwidth]{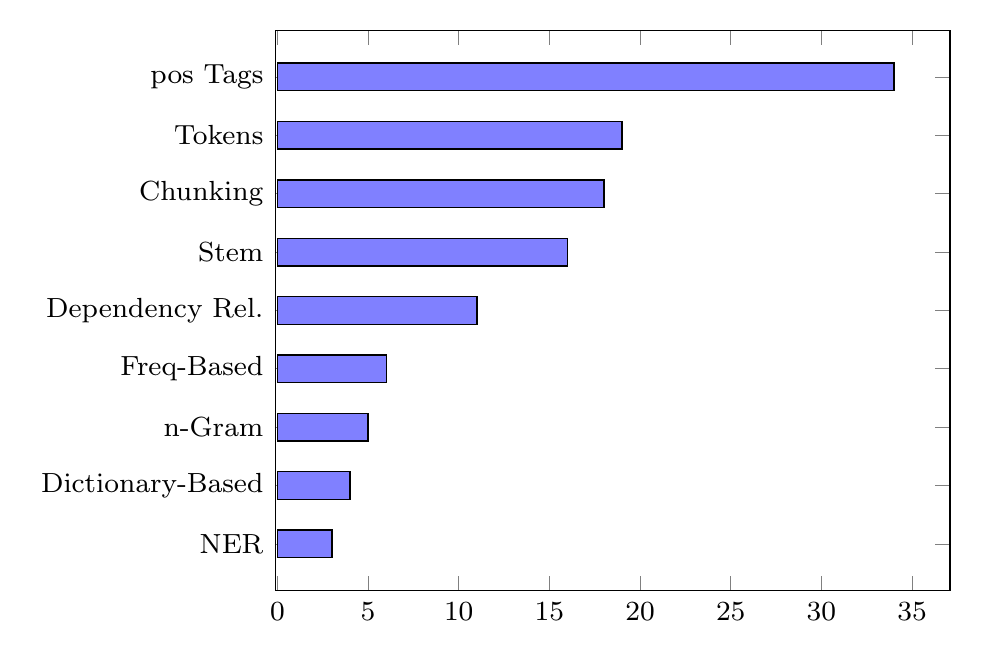}
     \caption{Frequently used features}
     \label{fig:freq_used_features}
 \end{figure}
 
 \begin{itemize}
     \item \textbf{The lexical and morphological features}, which are based on requirement words and their morphological information, such as token, stem (or lemma), n-gram (sequence of words).
     \item \textbf{The syntactic features}, which are derived from the syntax-related information in the requirement statement, such as POS-tag, chunks (noun or verb phrases), dependency relations, and the entities extracted using NER.
     \item \textbf{The frequency-based features}, which are based on the requirements list frequency metadata such as the number of words and the number of requirements.
     \item \textbf{Dictionary-based features}, which are extracted with the help of special dictionaries representing special words lists.
 \end{itemize}
 
  To further investigate the commonly used features, Fig. \ref{fig:freq_used_features} shows the most frequently used features and the number of papers that consider each of these features. The most used feature is the POS-tag, as it is considered in 34 out of the total 48 papers related to this type of representation.

 \subsubsection{Ontology-Based Representation} 

This group represents about 9\% of papers (9 out of 104). It benefits from pre-defined lexicons and semantic resources to extend the lexical and syntactic features. Most of these papers used a rule-based approach in their solutions by applying various ontology-path-based similarity measures. \cite{ferrari2012using,matsuoka2011ambiguity,arora2015change,ezzini2021using,baumer2018flexible}.

 \begin{figure}[h]
     \centering
     \includegraphics[width=0.6\textwidth]{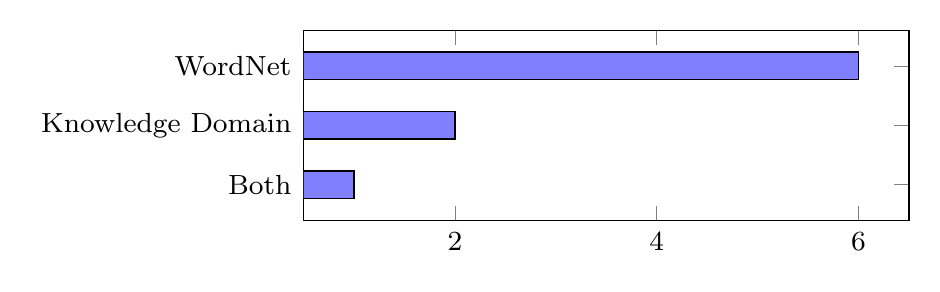}
     \caption{The used ontologies with their frequencies}
     \label{fig:ontologies}
 \end{figure}

One of the main differences between these papers is whether the used ontology is general or domain-specific. Fig. \ref{fig:ontologies} shows three types of works under this category based on the type of the used ontology: 6 papers used a general ontology (specifically WordNet) \cite{arora2016automated,gunecs2020automated,baumer2018flexible,matsuoka2011ambiguity,shah2021detecting,arora2015change}, while 2 papers used a domain-specific ontology \cite{ferrari2012using,ezzini2021using}, and one paper merges both approaches in its similarity calculations \cite{wang2016automatic}.

 \subsubsection{VSM}
\begin{figure}[H]
     \centering
     \includegraphics[width=0.6\textwidth]{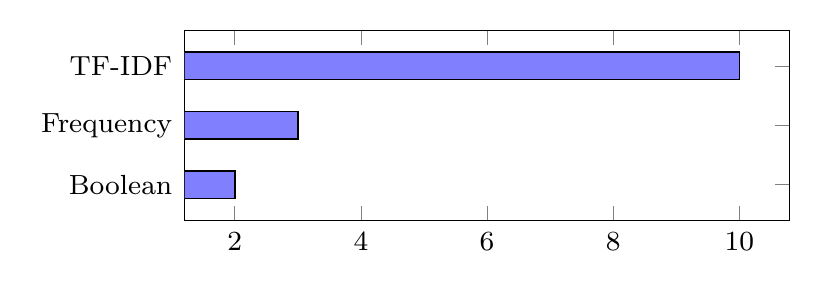}
     \caption{Weighting methods used in VSM representations with their frequency}
     \label{fig:vsm}
 \end{figure}

 Various forms of VSM representation are used in 15\% of papers (15 papers). These works combine the BOW technique with different weighting methods. 
 The most frequent combination is BOW with TF-IDF (10 papers). Fig \ref{fig:vsm} shows all used weighting methods with their frequencies.
 
 Based on this type of representation, each requirement is represented as a vector of words or tokens or stems combined with their weights. These vectors are used in the subsequent phase to apply similarity-based rules \cite{wang2021analyzing,ali2019exploiting} or as an input to train machine learning models (such as SVM \cite{jha2018using,rasekh2019mining,dias2020software,silva2020classifying}).
 
 \subsubsection{Topic modeling based representations}
\begin{figure}[h]
     \centering
     \includegraphics[width=0.6\textwidth]{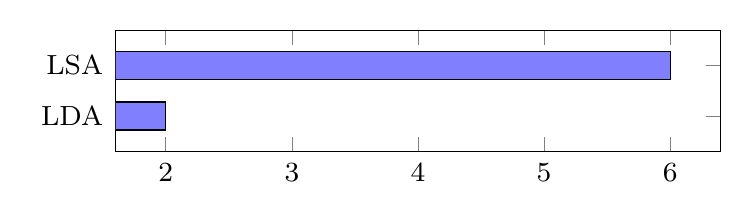}
     \caption{The used Topic modeling techniques with their frequency}
     \label{fig:topic_modeling}
 \end{figure}

 About 8\% of papers (8 papers) used topic modeling techniques to represent requirements texts based on automatically discovered latent topics. Two main topic modeling techniques are used in these papers: LSA and LDA. Fig \ref{fig:topic_modeling} shows the distribution of these two techniques over the related papers.

 The output of both techniques is a k-dimensional space that reflects a k latent topics within the processed requirements. Then, each requirement is represented as a vector representation in this k-dimensional space.
 In most related papers, this representation is employed to calculate the similarity between requirements as a part of similarity-based rules \cite{mahmoud2015role,misra2013entity} or clustering machine learning techniques \cite{misra2014latent,jiang2019recommending,misra2016topic}.

 \subsubsection{Advanced Embedding Representations}
 This part represents the second largest group of papers (24 papers) with a major increase in the last couple of years. It is noted that papers that use this kind of representations achieve promising results in many major tasks such as requirement classification \cite{hey2020norbert,palacio2019learning,kobilica2020automated}, traceability \cite{deshpande2021bert,das2021sentence,alhoshan2019using,sonbol2020towards}, ambiguity detection \cite{ferrari2019nlp}, and requirement extraction \cite{sainani2020extracting,wu2021identifying}.
 
\begin{figure}[]
     \centering
     \includegraphics[width=0.9\textwidth]{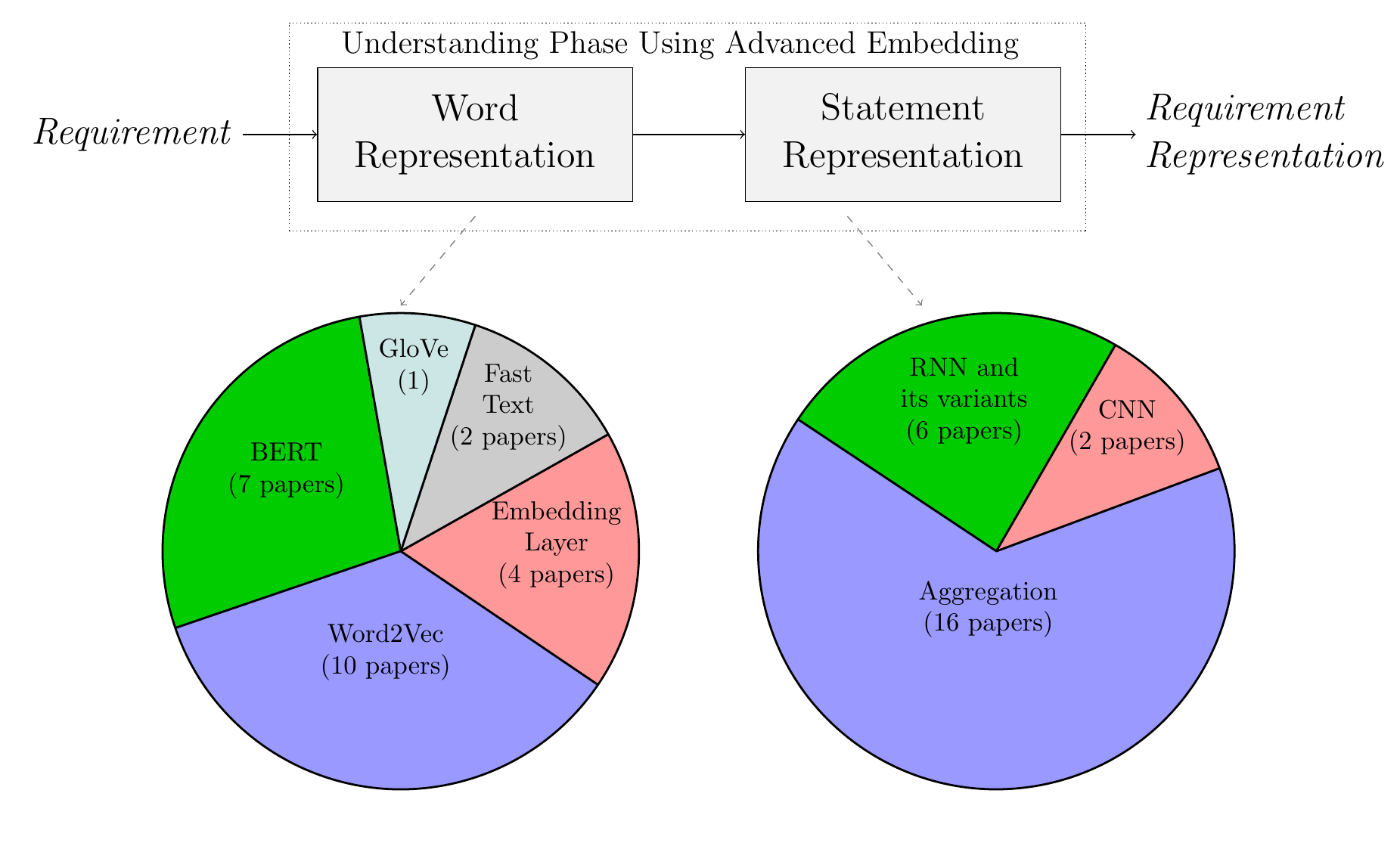}
     \caption{The general flow of representing requirements based on advanced embedding techniques}
     \label{fig:embedding}
 \end{figure}

 Based on this type of representation, each requirement is represented as a vector of floating point values. The proposed representations in the related papers are built following two generic steps (Fig. \ref{fig:embedding}): First, each token in the requirement text is represented using a word embedding technique; then the final requirement representation is constructed using the representation of its words. To further investigate papers that use advanced embedding techniques, we explore the techniques and models used in the implementation of each of these two steps.
 
 \vspace{5pt}
 
 \begin{itemize}
     \item \textbf{Word Representation:} Various word embedding techniques were used in the related literature. Word2vec and BERT are the most common techniques utilized in this group; together they were used in 17 out of the total 24 papers related to this group. Other works used GloVe and FastText models, while the remaining papers used an embedding layer instead (added to  the  front  of their neural network model).
     
     Table \ref{tab:word_embedding} shows the related papers for each of the used word embedding techniques.

     \begin{table}[h]
         \centering
         \begin{tabular}{p{0.24\linewidth} p{0.64\linewidth}}
              Word Embedding & Related Papers\\ \hline
              word2vec &  \cite{do2020capturing,mishra2019use,ferrari2019nlp,mishra2021crawling,alhoshan2019using,guo2017semantically,palacio2019learning,rahman2019classifying,sonbol2020towards,gulle2020topic}\\ 
              BERT & \cite{hey2020norbert,kicibert,das2021sentence,deshpande2021bert,wu2021identifying,de2021re,sainani2020extracting}\\ 
              Embedding Layer & \cite{baker2019automatic,kobilica2020automated,choetkiertikul2018deep,liu2021automated}\\
              GloVe & \cite{chatterjee2020identification}\\ 
              FastText & \cite{bhatia2020clustering,nicholson2021issue}\\
         \end{tabular}
         \caption{The used word embedding techniques with their related papers}
         \label{tab:word_embedding}
     \end{table}
     
     \item \textbf{Statement Representation:}
     Papers that use this type of representation can be further classified based on the method used to merge word embeddings in order to represent statements. Table \ref{tab:sentence_embedding} shows all related papers classified into three types of statement embedding techniques. 
     %\cite{ferrari2019nlp,mishra2021crawling}
     
     \begin{table}[h]
         \centering
         \begin{tabular}{p{0.28\linewidth} p{0.60\linewidth}}
            Statement embedding  &  Related Papers\\ \hline
            Aggregation-based  & \cite{hey2020norbert,kicibert,das2021sentence,deshpande2021bert,wu2021identifying,de2021re,sainani2020extracting,bhatia2020clustering,nicholson2021issue,do2020capturing,mishra2019use,alhoshan2019using,sonbol2020towards,gulle2020topic}\\
            
            RNN-based & \cite{chatterjee2020identification,kobilica2020automated,choetkiertikul2018deep,guo2017semantically,liu2021automated,rahman2019classifying}\\
            
            CNN-based & \cite{baker2019automatic,palacio2019learning}\\

         \end{tabular}
         \caption{the used statement embedding techniques with their related paper}
         \label{tab:sentence_embedding}
     \end{table}
     
     \vspace{5pt}
     
     \textit{The aggregation approach} is the most used technique (16 out of 24). It produces the requirement representation by applying various aggregation methods on word embedding results such as:
     \begin{itemize}
         \item Using the average of words embedding to represent statements \cite{do2020capturing,bhatia2020clustering,nicholson2021issue}.
         \item For BERT-based word embedding, a special token ([CLS]) is added to the beginning of the sentence and is used to calculate an aggregated representation of the statement \cite{hey2020norbert,deshpande2021bert,wu2021identifying,de2021re,sainani2020extracting}.
         \item Combining the syntactic information with the semantic ones by (1) classifying words based on their syntactic role, (2) calculating the weighted average for each role, then (3) concatenating the resultant vectors to form the final representation \cite{sonbol2020towards}.
         \item Representing requirements by representing its main "semantic frames" which can be retrieved with the help of FrameNet \cite{baker1998berkeley}. The representation of these frames is calculated by averaging the embedding vectors of their words \cite{alhoshan2019using}.
     \end{itemize}
     
     \vspace{5pt}
     
     \textit{Recurrent Neural Network (RNN) based representation} takes word embedding vectors as input. It finds a dense and low-dimensional semantic representation for each requirement statement by sequentially and recurrently processing its words. Many RNN architectures have been used in the related papers such as LSTM \cite{kobilica2020automated,choetkiertikul2018deep,rahman2019classifying}, Bi-LSTM \cite{chatterjee2020identification}, BI-GRU \cite{guo2017semantically}, and Skip-Thought \cite{liu2021automated}.
     
     \vspace{5pt}
     \textit{Convolutional Neural Network (CNN) based representation} takes word embedding vectors as input. It finds the final representation through a number of convolutional layers, pooling layers, and fully connected layers \cite{baker2019automatic,palacio2019learning}.

 \end{itemize}

\subsection{(RQ4) What are the main research directions to represent requirements for each category of RE tasks?}

To answer this question, we start with a general overview of how the research directions to represent requirements have evolved over the last decade; then we present a deeper analysis of the trends and possible future directions in each category of RE tasks.

\subsubsection{General Overview}
\begin{figure}[h]
 \centering

 \includegraphics[width=0.6\textwidth]{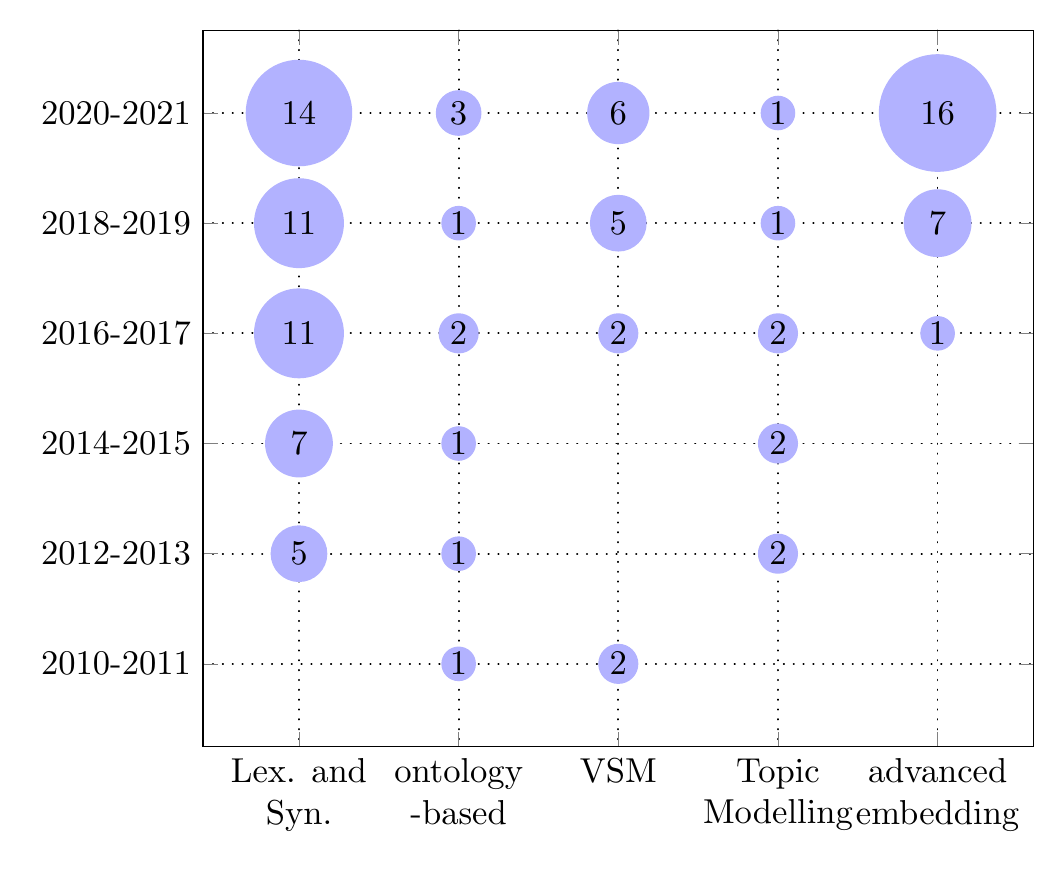}
 
 \caption{Bubble chart showing the number of papers according to the used NLP representation and the year of publications}
 \label{fig:bubble_chart_NLP_year}
\end{figure}

Fig. \ref{fig:bubble_chart_NLP_year} shows the number of papers according to the used NLP representation and the year of publication. The bubble chart (Fig. \ref{fig:bubble_chart_NLP_year}) shows clearly how the trend has changed in the last few years from the lexical and syntactic features to the advanced embedding techniques. These embedding representations became the most used type in the last 2 years (2020-2021) with 15 out of 40 published works in that period.

\subsubsection{Research directions for each category of RE tasks}

To further investigate the trends of various representations, we studied their distribution over the main categories.

\begin{figure}[h]
 \centering

 \includegraphics[width=0.6\textwidth]{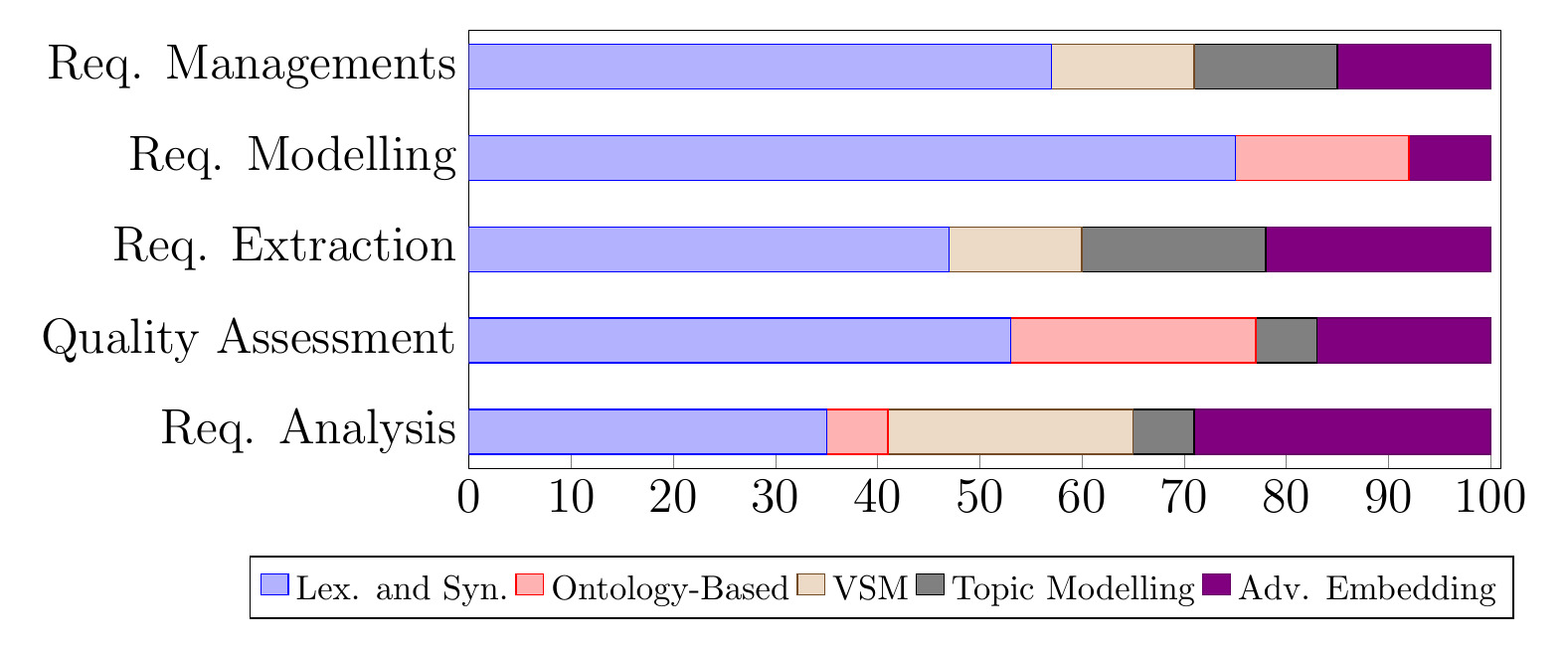}
 
 \caption{The percentage of used NLP representation for each category of RE tasks}
 \label{fig:nlp_per_RE_task}
\end{figure}

Fig. \ref{fig:nlp_per_RE_task} shows the number of papers that used each of these representations for each category. As noted, representing requirements based on a set of lexical and syntactic features is the most used approach in many tasks, especially in quality assessment and requirement modeling. However, there is a clear interest in the use of advanced embedding and VSM-based representations in requirement analysis tasks.

The following sub-sections discuss the trends for each category.

\begin{itemize}
\vspace{5pt}
\item \textbf{Requirement Analysis:}
Embedding techniques have been proven to be useful for various requirement analysis tasks \cite{hey2020norbert,deshpande2021bert,sonbol2020towards,alhoshan2019using}. Two recent studies \cite{hey2020norbert, deshpande2021bert} compared the performance of lexical-syntactic features representation with advanced embedding representation, and concluded that using embedding vectors led to better results in requirement classification \cite{hey2020norbert} and traceability detection \cite{deshpande2021bert} tasks. This finding is consistent with the results observed in the NLP literature \cite{li2018word}, which indicate that embedding techniques are powerful in determining the similarity of words and sentences.
of words and sentences. Similarity represents the core operation in almost all requirements analysis tasks.

 \vspace{5pt}
 
\item \textbf{Quality Assessment:}
More than 50\% of quality assessment papers used lexical and syntactic features to represent requirements (Fig. \ref{fig:nlp_per_RE_task}).
However, a deeper analysis of related papers showed that we can differentiate between two types of quality-related RE tasks:
\begin{itemize}
    \item The majority of papers focused on lexical and syntactic quality-related issues such as lexical ambiguity and conformance with templates. Many significant contributions provide experiments that indicate the effectiveness of using lexical and syntactic features to represent requirements when handling this category of tasks \cite{ferrari2018detecting,femmer2017rapid,arora2015automated,dalpiaz2019detecting}.
    \item The remaining papers focused on semantic quality-related issues such as incompleteness, and semantic-level quality problems. For these tasks, the most used representation in the literature is ontology-based representation \cite{ezzini2021using,baumer2018flexible,ferrari2014measuring}. Besides, one of the recent interesting works employed advanced embedding representation based on an embedding layer with a Skip-Thought model  to handle incompleteness detection problem \cite{liu2021automated}.
\end{itemize}  

 \vspace{5pt}
\item \textbf{Requirements Extraction:}
It is noted that there are two directions to handle requirements extraction tasks based on the source of data:
\begin{itemize}
    \item Extracting requirements from documents: the structured format of these documents motivated many researchers to use lexical and syntactic features. All retrieved papers under this sub-category used lexical and syntactic representations \cite{shi2021automatically,abualhaija2020automated,haris2020automated}.
    
    \item Extracting requirements from reviews, forums, or similar applications: Most of the recent papers (especially the ones published in the last 2 years) used advanced embedding techniques to handle these tasks \cite{de2021re,wu2021identifying,do2020capturing}. This finding could be due to the approaches used to handle this type of requirement extraction tasks; most solutions handle the problem as a classification task (classify sentences or phrases into requirements or not requirements classes) or a clustering task (group sentences that mention the same aspects in the product). Both approaches (clustering and classification) make the embedding technique more suitable (as mentioned previously when discussing the requirement analysis case).
\end{itemize}

 \vspace{5pt}
\item \textbf{Requirements Modeling:}
Almost all works under this category represent requirements based on their lexical and syntactic features or using ontology-based representation \cite{haj2021semantic,tiwari2019approach,gunecs2020automated}. These features are used in the form of rules and syntactic patterns to generate various models based on rule-based transfer techniques.

 \vspace{5pt}
\item \textbf{Requirements Management:}
Most of the papers under this category use lexical and syntactic features representation \cite{shafiq2021nlp4ip,kifetew2021automating,mczara2015software}. However, the number of NLP-based publications related to this category is still limited to have a clear conclusion regarding the suitable representations for this type of tasks.
\end{itemize}

\subsection{(RQ5) What gaps and potential future directions exist in this field?}

Our survey shows that the use of NLP-based requirements representation has made remarkable progress over the last decade. Based on our findings in the previous research questions (especially RQ4), the use of advanced embedding techniques to represent requirements seems to be promising for most RE tasks (mainly requirement analysis, extracting requirements from reviews and forums, and semantic-level quality tasks). Besides, representing requirements based on a set of lexical and syntactic features or using ontology-based information seems more suitable for other RE tasks (such as modeling and syntax-level quality tasks).

In the following paragraphs, we will discuss the gaps that represent potential future research directions in this domain.

\subsubsection{More RE domain-specific word embedding models}
As shown in Fig. \ref{fig:embedding}, word representation plays an essential role in representing requirements (mainly when using embedding techniques). To implement this step, most researchers use generic word embedding pre-trained models in their solutions. One of the main limitations of these generic models is their inability to provide appropriate representations for many domain-specific words. For example, the domain-specific meaning of words like "virus", "cookies", "Python", "fork", etc. can not be captured based on models trained on generic corpora \cite{mishra2021crawling}. Building such domain-specific word embedding models is notoriously a challenging task, especially with the little size of data available in the RE domain. One of the recent works \cite{mishra2021crawling} provided an embedding model trained on 92 MB of texts collected from Wikipedia pages related to the software engineering domain. However, more research is needed (1) to have embedding models trained on more practical industrial texts, (2) and to evaluate the use of these models in various RE tasks.

\subsubsection{More suitable statements representation techniques}

Another important challenge, related to embedding techniques, is the way of merging word vectors to formulate requirements representations (step 2 in Fig \ref{fig:embedding}). Although the suggested techniques lead to good accuracy in some RE tasks (such as requirements classification and traceability detection), these representations can still be considered as a step towards more suitable representations. Most of these works use aggregation techniques (that ignore word order \cite{de2016representation}) or RNN-based technique (that usually focus on predicting the next element in a sequence \cite{dupond2019thorough}). These works can be further improved by considering more RE domain-specific representations \cite{sonbol2020towards}. One of the main potential future directions to handle this problem is using a "syntax-aware sentence embedding" technique that considers both the main syntactic roles in requirement statements (such as actor, action, objects,...) and the semantic aspects of requirements.

\subsubsection{More adaptive syntactic processing}
Many studies have concluded that there is a gap between RE tasks automation research and its implementation undertaken in industrial and real-life projects \cite{liu2021requirements,mavin2017does, eklund2014industrial}. This gap is more obvious in the case of rule-based approaches since they usually need the requirements to be represented based on specific templates \cite{fanmuy2012requirements}; hence, their success strongly depends on the consistency of the requirements with the predefined templates \cite{arora2015automated,fanmuy2012requirements}. Using such “template-based” approaches to handle RE tasks could lead to lower accuracy when applied to new requirements written based on some variations of the predefined template or a completely different template. These situations are common in real-life projects when it is hard to control requirements authoring environments, especially in large development projects, or when one has little control over the requirements authoring environments. \cite{arora2015automated,wautelet2014unifying,liu2021requirements,mavin2017does,eklund2014industrial,fanmuy2012requirements,ferrari2017natural}.
One of the potential future research directions is working on more adaptive approaches which can be more flexible when handling requirements, such as identifying the syntactic structures automatically and building more adaptive approaches based on the dynamically identified structure.

\subsubsection{More research in some RE tasks}
Only a few works handled the following RE tasks: requirements prioritization, effort estimation, and semantic and pragmatic level quality tasks. More research should be conducted to explore the efficiency of using advanced embedding techniques for these tasks.

\section{Main Findings:}

In this section, we summarize the main conclusions and findings in the following points:
\vspace{3pt}
\begin{enumerate}[(1)]
    \item There is a significant increase in the number of papers that use NLP-based requirements representation over the last decade. The increase is more obvious in the last 2 years.
    \vspace{3pt}
    \item We recognized 104 papers that employed an NLP-based representation to solve various RE tasks. We classified these works based on (1) the targeted RE task using a hierarchy consisting of 6 top-level categories  and 19 sub-categories (Fig. \ref{fig:treemap}), (2) the used NLP-based representation using 5 classes of text representation techniques (Fig. \ref{fig:general_flow}).
    \vspace{3pt}
    \item Using advanced embedding techniques to represent requirements seems to be promising for most RE tasks (especially requirement analysis, extracting requirements from reviews and forums, semantic-level quality tasks). Besides, representing requirements based on a set of lexical and syntactic features or using ontology-based information seems more suitable for other RE tasks (such as modeling and syntax-level quality tasks).
    \vspace{3pt}
    \item Most of the proposed embedding-based representations use generic language models that suffer from limitations when representing the meaning of software engineering terms. Having more domain-specific language models to represent words is considered one of the important research directions in future works.
    \vspace{3pt}
    \item Most of the proposed embedding-based techniques represent requirements statements without giving attention to the special structure of a software requirement that consists of clear components (actor, action, data object, etc). Using syntax-aware statements embedding could be one of the possible solutions to have more domain-specific representations that reflect the semantics of requirements in a better way (compared to traditional aggregation approaches).
    \vspace{3pt}
    \item There is a need for more flexible approaches in terms of the ability to handle requirements that do not follow (completely or partially) standard requirements templates, which is the common case in real-life projects.
    \vspace{3pt}
    \item The number of researches that use NLP-based representation is still limited for some RE tasks (requirements prioritization, effort estimation, semantic-based quality assessment). 
\end{enumerate}

Finally,  tables \ref{tab:t_all_papers} and \ref{tab:t3} summarize the extracted information for all 104 selected papers, in addition to their categorization based on their targeted RE task and the used text representation.
In the matrix represented in Table \ref{tab:t_all_papers}, each cell includes the papers related to each category of RE tasks and text representation. Note that two papers (\cite{dollmann2016and} and \cite{sainani2020extracting}) are mentioned twice in the table since they are related to both requirement analysis and requirement extraction categories.
Table \ref{tab:t3} shows a detailed summary of the 104 papers. This table includes detailed information about the input and the output of the approach proposed in each paper. In addition, it encloses other descriptive information including the employed text processing steps, details about the solution technique (e.g., which ML approach is used), the used dataset,  and a summary of the results. This table provides a starting point for researchers and practitioners in this field to obtain a quick
overview of the state-of-the-art in each RE task, including the recommended dataset to be used and the most related publications.

\begin{table*}
    \begin{tabular}{p{0.14\linewidth} p{0.14\linewidth} p{0.14\linewidth} p{0.14\linewidth} p{0.14\linewidth} p{0.14\linewidth}}
         & Lexical and  \newline Syntactic Features 
         & Ontology-Based 
         & VSM 
         & Topic  \newline Modeling-Based 
         & Advanced Embedding
         \\
         \hline
         Requirement \newline Analysis
         &
         \cite{di2013automatic,rashwan2013ontology,aldekhail2022intelligent,dollmann2016and,ott2013automatic,mahmoud2016detecting,dalpiaz2019requirements,wang2015semantic,mezghani2018industrial,deshpande2019data,blake2015shared,li2020ontology,samer2019new,younas2020extraction,casamayor2012functional,diamantopoulos2017software,leitao2021srxcrm}
         &\cite{arora2015change,shah2021detecting,wang2016automatic}
         &\cite{casamayor2010identification,wang2021analyzing,raharja2019classification,dias2020software,kobilica2020automated,eyal2018semantic,ali2019exploiting,chhabra2017requirements,silva2020classifying,rasekh2019mining,khatian2021comparative,krasniqi2021analyzing,sultanov2011application} 
         &\cite{wang2020automated,misra2016topic,mahmoud2015role}
         &\cite{baker2019automatic,palacio2019learning,sainani2020extracting,rahman2019classifying,gulle2020topic,deshpande2021bert,sonbol2020towards,hey2020norbert,guo2017semantically,nicholson2021issue,kicibert,chatterjee2020identification,das2021sentence,alhoshan2019using} 
         \\
         \hline
         
         Requirements \newline Extraction
         &\cite{peng2016approach,abualhaija2020automated,quirchmayr2018semi,wang2016experiment,yang2021phrase,haris2020automated,li2018automatically,shi2021automatically,dollmann2016and}
         & 
         &\cite{jha2018using,abbas2020automated} 
         &\cite{jiang2019recommending,bakar2016extracting,carreno2013analysis}
         &\cite{lu2017automatic,wu2021identifying,do2020capturing,de2021re,sainani2020extracting} 
         \\
         \hline

         Quality \newline Assessment
         &\cite{ferrari2018detecting,cruz2017detecting,arora2015automated,parra2015methodology,dalpiaz2019detecting,wang2013automatic,osama2020score,ferrari2014measuring,femmer2017rapid}
         &\cite{baumer2018flexible,matsuoka2011ambiguity,ferrari2012using,ezzini2021using}
         & 
         &\cite{misra2013entity}
         &\cite{liu2021automated,ferrari2019nlp,mishra2019use} 
         \\
         \hline

         Modeling
         &\cite{sree2018extracting,haj2021semantic,thakur2016identifying,tiwari2019approach,hamza2019generating,hamza2015recommending,al2018use,elallaoui2018automatic,lucassen2017extracting}
         &\cite{gunecs2020automated,arora2016automated}
         & 
         & 
         &\cite{bhatia2020clustering} 
         \\
         \hline

         Management
         &\cite{hussain2013approximation,mczara2015software,shafiq2021nlp4ip,kifetew2021automating}
         & 
         &\cite{ali2021requirement} 
         &\cite{misra2014latent}
         &\cite{choetkiertikul2018deep} 
         \\
         \hline

         Others
         & 
         & 
         & 
         & 
         &\cite{mishra2021crawling} 
         \\
         \hline
    \end{tabular}
    \caption{The final list of selected papers categorized based on the type of targeted RE task and the used text representation}
    \label{tab:t_all_papers}
\end{table*}

\section{Threats To Validity}
Likewise any secondary research process, it is almost impossible to guarantee that we found the entire population of all the relevant papers. However, several actions were undertaken to minimize threats to validity.

\begin{itemize}
    \item To ensure the inclusion of almost all relevant academic works in the field, we followed a systematic mapping review methodology based on the recommenced guidelines for similar cases.\cite{petersen2015guidelines}.
    \item Five reputable and well-known data sources ( Scopus, IEEE Xplore, ACM Digital Library, ScienceDirect, and SpringerLink) were chosen to maximize the number of candidate papers.
    \item We tried to make our search string as general as possible by including various synonyms for each term and by including papers that cover either syntactic or semantic aspects. However, the final search string may not encompass all the existing synonyms, which might lead to not capturing all the relevant studies. We mitigated this threat by checking the references of the final selected papers to add any additional relevant works.
    \item To minimize mistakes caused by subjective analyses, we followed a rigorous study selection process, guided by clear inclusion and exclusion criteria. However, the exclusion of papers published in languages other than English may have failed to potentially find some relevant works.
    \item When there were doubts or conflicts about whether to include an article or not, the final decision is discussed between authors.
    \item To obtain data consistency and avoid bias in data extraction, we defined a clear data extraction template and  discussed our results in several brainstorming sessions.
\end{itemize}

\section{Conclusion}
This study presents a systematic mapping review of the used NLP-based representations in various RE tasks. Starting from 2,227 papers retrieved from five well-known digital libraries, we recognized 104 primary papers fulfilling the inclusion and exclusion criteria. We analyzed these works and categorized them based on: the targeted RE task and the used text representation.

Our results indicate that about two-thirds of retrieved publications handle tasks related to requirements classification, requirements traceability, ambiguity detection, and extracting requirements from reviews and documents. On the other hand, Lexical and syntactic features are widely used to represent requirements (more than 45\% of publications). Besides, a growing number of papers use advanced word embedding techniques, especially in the last two years.
Moreover, we summarize the main research directions to represent requirements in each category, and identify the gaps and possible future directions.

The data extracted from the selected 104 papers and their categorization, can help as a useful set of references for further analysis in each task or category of tasks in RE. Besides, trends and gaps identified from this mapping study have provided many new ideas for research opportunities.

\begin{landscape}
    \fontsize{7.5}{10}\selectfont
    \begin{longtable}
    {p{0.08\linewidth} | p{0.07\linewidth} | p{0.05\linewidth} | p{0.06\linewidth}  | p{0.08\linewidth}| p{0.055\linewidth}| p{0.18\linewidth}| p{0.11\linewidth}| p{0.07\linewidth}| p{0.08\linewidth}}
   \textbf{Paper} 
    & \textbf{Targeted Task}
    & \textbf{Input}
    & \textbf{Output}
    & \textbf{Representation Method}
    & \textbf{Solving Method}
    & \textbf{Text Processing}
    & \textbf{Solving Method Details}
    & \textbf{Dataset}
    & \textbf{Results}
    \\
    \hline
  Baker et al. \cite{baker2019automatic}
    & Classification
    & a Req.
    & NFR classes
    & Advanced Embedding
    & ML
    & Stop words removal; Tokenization; Stemming;
    & ANN; CNN*
    & PROMISE+
    & F1: 0.82-0.92
    \\
    \hline
  Dias et al.\cite{dias2020software}
    & Classification
    & a Req.
    & FR/NFR; NFR classes
    & VSM
    & ML
    & Stop words removal; Tokenization; Stemming; BOW, TFIDE, CH2 feature selection
    & SVM*; MNB; KNN; LR
    & PROMISE (lima version)
    & F1: FR/NFR 0.91 / NFR: 0.72
    \\
    \hline
    Casamayor et al. \cite{casamayor2010identification}
    & Classification
    & a Req.
    & NFR classes
    & VSM
    & ML
    & Sop words removal; Stemming; BOW + TF-IDF
    & Bayesian Classifier +EM strategy
    & PROMISE
    & Acc: 0.75
    \\
    \hline
    Dalpiaz et al. \cite{dalpiaz2019requirements}
    & Classification
    & a Req.
    & FR/QR
    & Lexical-Syntactic features
    & ML
    & 17 lexical-syntactic features including syn. dependency-based features
    & SVM
    & PROMISE+
    & F1: F(0.79); Q(0.76)
    \\
    \hline
  Hey at al. \cite{hey2020norbert}
    & Classification
    & a Req.
    & FR/QR; NFR classes; FR classes
    & Advanced Embedding
    & ML
    & BERT
    & FFNN
    & PROMISE (Dalpiaz Version)
    & FR/QR up to 0.94; NFR 0.87; FR up to 0.92
    \\
    \hline
    Khatian et al. \cite{khatian2021comparative}
    & Classification
    & a Req.
    & NFR classes
    & VSM	
    & ML	
    & Stop words removal; Stemming; BOW	
    & DT; KNN; RF; NB; LR*
    & PROMISE
    & F1: 0.75
    \\
    \hline
  Rahman et al. \cite{rahman2019classifying}
    & Classification
    & a Req.
    & NFR classes	
    & Advanced Embedding	
    & ML	
    & word2vec
    & LSTM; GRU; CNN
    & PROMISE
    & F1: 0.71
    \\
    \hline
  Rashwan et al. \cite{rashwan2013ontology}
    & Classification
    & a Req.
    & NFR classes
    & Lexical and Syntactic Features	
    & ML	
    & Tokenization, Sentence splitter, Stemming;
    &  	SVM
    & CONCORDIA +PROMISE
    & F1: 0.67-0.84
    \\
    \hline
    Younas et al. \cite{younas2020extraction}
    & Classification
    & a Req.
    & NFR classes
    & Lexical and Syntactic Features
    & 	Rule-Based
    & 	stopwords removal, stemming, POS-tagging, famous NFR indicator keywords
    & 	similarity-based approach
    & 	PROMISE	
    & F1: 0.64
    \\
    \hline
    Mahmoud et al. \cite{mahmoud2016detecting}
    & Classification
    & a Req.
    & NFR classes
    & lexical and syntactic features
    & Rule-Based
    & lemmatization; Normalized Google Distance (NGD)
    & similarity-based approach
    & Own Dataset
    & R: 0.88 /P: 0.52
    \\
    \hline
   Kici et al. \cite{kicibert}
    & Classification
    & a Req.
    & FR/NFR
    & Advanced Embedding
    & ML
    & word embedding (DistillBERT and BiLSTM + ELMo)	
    & multi-class text classification
    & DOORS + PROMISE	
    & F1: 0.80
    \\
    \hline
   Raharja et al. \cite{raharja2019classification}
    & Classification
    & a Req.
    & Quality aspects
    & VSM	
    & ML	
    & Tokenization, Stop words removal, Stemming, TF-IDF	
    & Fuzzy similarity measure + KNN	
    & 6 datasets (PROMISE+)
    & P:0.68; R:0.55
    \\
    \hline
   Chatterjee et al. \cite{chatterjee2020identification}
    & Classification
    & a Req.
    & Significant FR.
    & Advanced Embedding
    & ML
    & GloVe trained on 124 SRS
    & Bi-LSTM + Attention
    & Own Dataset
    & F1: 0.86
    \\
    \hline
  Palacio et al. \cite{palacio2019learning}
    & Classification
    & a Req.
    & Sec/NSec
    & Advanced Embedding
    & ML
    & word2vec trained on a collected security dataset
    & CNN
    & Own Dataset
    & Acc: 0.71-0.96
    \\
    \hline
   Kobilica  et al.\cite{kobilica2020automated}
    & Classification
    & a Req.
    & Sec/NSec	
    &VSM	
    &ML	
    & BOW
    &22 ML algorithm (including LSTM*)
    &SecReq
    &Acc: 0.84
    \\
    \hline
   Li et al.\cite{li2020ontology}
    & Classification
    & a Req.
    & Sec/NSec	
    & Lexical and Syntactic Features	
    & ML	
    & 140 Keywords (Lexical features) + 34 linguistic rules (Syntactic Features)
    & 	J48; NB; LR	
    & SecReq +PROMISE
    & F1: 0.63
    \\
    \hline
    Silva et al. \cite{silva2020classifying}
    & Classification
    & a Req.
    & Design Patterns
    & VSM	
    & ML
    & TF-IDF
    & LR, MNB, SVM*, RF
    & PROMISE+	
    & F1: 0.52
    \\
    \hline
   Ott at al. \cite{ott2013automatic}
    & Classification
    & a Req.
    & Req. Topic	
    & Lexical and Syntactic Features
    & 	ML	
    & Tokenization, n-gram, Surrounding requirements
    & 	MNB, SVM*
    & 	DCU
    & 	R:0.8/  P: 0.6
    \\
    \hline
   Krasniqi et al. \cite{krasniqi2021analyzing}
    & Classification
    & Issues
    & quality related issue
    & VSM
    & ML	
    & Length-based features, BOW, n-gram, syntactic features, Semantic Triplet	
    & RF	
    & Own Dataset
    & F1:0.79
    \\
    \hline
   Gulle et al. \cite{gulle2020topic}
    & Clustering
    & a set of Req.
    & Clusters
    & advanced embedding
    & ML
    & Tokenization; Stop words removal; Remove Template Words; word2vec
    & Word Mover’s Distance based clustering
    & 	CrowdRE
    & 	no measure exists
    \\
     \hline
   Casamayor et al. \cite{casamayor2012functional}
    & Clustering
    & use-cases
    & Clusters for functional group
    & lexical and syntactic features
    & ML	
    & POS-tagging; chunking;  categorize candidate responsibilities based on the NP and V relates to.
    & 	Clustering Based on EM, CobWeb, Xmeans, DBScan
    & 	Own Dataset	
    & Rand Index: 0.80   
    \\
     \hline
    Misra et al. \cite{misra2016topic}
    & Clustering
    & SRS Document
    & Clusters
    & Topic modeling
    & ML	
    & pos tagging; chunking; Remove stop-words; lemmetization; LSA
    & Theme based clustering
    & 	Own Dataset	
    & Purity: 0.46; F1: 0.52. 
    \\
     \hline
    Eyal et al. \cite{eyal2018semantic}
    & Clustering
    & a set of Req.	
    & Clusters
    & VSM
    & ML
    & Tokenization; stopwords removal; stemming; VSM
    & Agglomerative Hierarchical clustering (AHC)
    &  Own Dataset
    & P:0.72-0.83/ R:0.54-0.61 
    \\
     \hline
    Mezghani et al. \cite{mezghani2018industrial}
    & Clustering
    & SRS Document
    & Clusters
    & lexical and syntactic features
    & ML
    & pos tagging; NP chunking to detect technical business terms
    & 	K-mean
    & 	Own Dataset
    & 	 No Available Results
    \\
    \hline 
   Das et al. \cite{das2021sentence}
    &Traceability
    &Two Reqs
    &Similarity score
    &Advanced Embedding
    &Rule-Based
    &BERT
    &Cos Similarity
    &800 pairs of reqs
    &Acc: 0.88
   \\
    \hline 
    Blake et al. \cite{blake2015shared}
    &Traceability
    &Two SRS documents
    &identify overlapping requirements
    &lexical and syntactic features
    &Rule-Based
    &Tokenization; stopwords removal; generate synonyms; determine verbs
    &Similarity between reqs is calculated based on number of similar words 
    &Own Dataset
    & Not Available
    \\
    \hline 
    Sonbol et al. \cite{sonbol2020towards}
    &Traceability
    &Two Reqs
    &similarity score
    &Advanced Embedding
    &Rule-Based
    &Tokenization; POS Tagging; NP chunking; classify requirements to semantic dimensions; word2vec
    &Manhattan Distance Based Similarity
    &5,852 pairs of reqs
    &F1: 0.92
    \\
    \hline 
    Alhoshan et al. \cite{alhoshan2019using}
    &Traceability
    &Two Reqs
    &Similarity score
    &Advanced Embedding
    &Rule-Based
    &tokenisation,  stop-word  removal, POS tagging  and  lemmatisation, Embedding-based  representation  of  semantic  frames
    &relatedness score based on frame Embeddings and cos sim
    &1,770 pairs of reqs
    &F1: 0.86
    \\
    \hline 
    Deshpande et al. \cite{deshpande2019data}
    &Traceability
    &Two Reqs
    &Dependency and its type
    &lexical and syntactic features
    &ML
    &tokenized, stop words removal; lemmatization
    &RF, SVM, NB
    &PURE dataset
    &F1:0.89
    \\
    \hline 
    Samer et al. \cite{samer2019new}
    &Traceability
    &Two Reqs
    &related / not related
    &lexical and syntactic features
    &ML
    &Stop words Removal , merged synonyms, lemmatization
    &Linear SVM, NB,RF*, and KNN
    &OpenReq Dataset
    &F1: 0.89
    \\
    \hline 
    Sultanov et al. \cite{sultanov2011application}
    &Traceability
    &Two Reqs
    &related / not related
    &VSM
    &ML
    &stop words removal, stemming, VSM+TF-IDF; agent words
    &swarm intelligence
    &Own Dataset
    &F1: 0.58
   \\
    \hline 
    Lu et al. \cite{lu2017automatic}
    &Traceability
    &Two Reqs
    &related / not related
    &lexical and syntactic features
    &rule based
    &(A) Jaccard sim based approach
    (B) stop words removal; stemming; SVM+Freq; Cos similarity
    &apply two approaches and use fuzzy logic to merge results
    &Own Data from 14 projects
    &Acc: 0.83
    \\
    \hline 
    Deshpande et al. \cite{deshpande2021bert}
    &Traceability
    &two Reqs
    &related / not related
    &advanced embedding
    &ML
    &stop words removal; tokenization; lemmatization; BERT
    &BertForSequence Classification
    &Redmine Dataset
    &F1: 0.93
   \\
    \hline 
    Guo et al. \cite{guo2017semantically}
    &Traceability
    &SRSs and Design documents
    &related / not related
    &advanced embedding
    &ML
    &word2vec trained on a domain corpus
    &RNN (Bidirectional Recurrent Gated Unit)
    &Own Dataset
    &Mean Average Precision: 0.59
   \\
    \hline 
   Chhabra et al. \cite{chhabra2017requirements}
    &Traceability
    &use cases+code
    &code-reqs links
    &VSM
    &rule based
    &VSM based model
    &a set of huristics
    &Own Dataset
    &F1: 0.56
    \\
    \hline 
  Wang et al. \cite{wang2021analyzing}
  &Traceability
 &reqs and artifcats 
 &trace links
 &VSM
 &rule based
 &normalization, stemming, VSM, word2vec
 &ranking technique
 &Easyclinic, iTrust, eTour
 & Not Available
    \\
    \hline 
  Wang et al. \cite{wang2020automated}
 &Traceability
 &reqs and artifcats 
 &trace links
 &Topic modeling
 &ML
 &stop words removal, pos tagging, stemming, TF–IDF, Topic Modeling
 &Biterm Topic Model–Genetic Algo
 &Own Dataset
 &R/P larger than 0.70/0.30
    \\
    \hline
 Rasekh et al. \cite{rasekh2019mining}
 &Traceability
 &reqs and artifcats 
 &trace links
 &VSM
 &ML
 &TF-IDF+BoW
 &Bayesian learning , RBF Network, LR, SVM, DT
 &Easyclinic, albergate
 &P:0.91; R:0.94
    \\
    \hline
    Mahmoud et al. \cite{mahmoud2015role}
    &Traceability
    &reqs and artifcats 
    &trace links
    &Topic modeling
    &rule based
    &VSM, VSM with thesaurus support (VSM-T), POS-enabled VSM (VSM-POS), latent semantic indexing (LSI), LDA, explicit semantic analysis (ESA), and normalized Google distance (NGD).
    &similarity function
    &CM-1, eTour, and iTrust.
    &VSM-T, VSM-POS, ESA, and NGD outperform LSA and LDA.
    \\
    \hline
    Alia et al. \cite{ali2019exploiting}
    &Traceability
 &reqs and artifcats 
 &trace links
 &VSM
 &rule based
 &pos tagging, stemming, 
 &apply baseline approaches + Constraint-based pruning to approve links
 &iTrust, Pooka, Lynx, SIP
 &11\%-107\% and 8\%-64\% higher P and R than VSM and JSM
    \\
    \hline
    Aldekhail et al. \cite{aldekhail2022intelligent}
    &Traceability
 &a set of reqs
 &conflicts
 &lexical and syntactic features
 &rule based
 &basic NLP processing
 &rule based system to identify conflicts
 &CPMS, ENP, FMS
 &Accuracy: 100\%
    \\
    \hline
 Shah et al. \cite{shah2021detecting}
 &Traceability
 &a set of reqs
 &conflicts
 &ontology-based
 &rule based
 &tokenization, pos tagging, parser, dependency tree, WordNet
 &clustering requirements, then process clusters to detect conflicts
 &5 datasets
 &F1:0.65
    \\
    \hline
 Arora et al. \cite{arora2015change}
 &Traceability
 &set of reqs+change senario
 &inter-requirement dependency
 &ontology-based
 &rule based
 &chunking; extract NP and VP; similarity measure using Levenstein and path based sim
 &similarity based algorithm
 &Two industrial case studies
 &detected 99\% of the impacted reqs
    \\
    \hline
    Nicholson et al. \cite{nicholson2021issue}
 &Traceability
 &a set of reqs
 &dependencies
 &advanced embedding
 &ML
 &word embedding
 &LR*, RF, NN
 &3 test cases
 &F1: 0.55-0.67
    \\
    \hline
Leitao et al. \cite{leitao2021srxcrm}
 &Traceability
 &Design Spec. and Reqs
 &Association Rules
 &lexical and syntactic features
 &ML
 &tokenization, pos tagging, chunking
 &Association Rule Mining
 &Own Dataset
 &238 rules were extracted
    \\
    \hline
    Diamantopoulos et al. \cite{diamantopoulos2017software}
    &Semantic Role Labeling
    &a Req
    &Concepts + Relations
 &lexical and syntactic features
 &ML
 &Fetures: lemma, pos, syntactic dependency relations
 &logistic regression classifier
 &Own Data
 &F1: 0.76
    \\
    \hline
    Wang et al. \cite{wang2016automatic}
 &Semantic Role Labeling
 &a Req
 &semantic roles
 &ontology-based
 &ML
 &Fetures: lemmas; pos tags;  chunking; Verb class; parsing tree; word sense fetures using wordNet and domain ontology
 &The maximum entropy classifier
 &Own Dataset
 &F1:0.85
    \\
    \hline
    Wang et al. \cite{wang2015semantic}
 &Semantic Role Labeling
 &SRS
 &semantic roles
 &lexical and syntactic features
 &ML
 &tokenization, pos tagging, parsing, NER
 &DT classifier
 &18 SRS
 &P:  0.92-0.93; R: 0.91-0.92
    \\
    \hline
Ezzini et al. \cite{ezzini2021using}
 &Ambiguity Detection
 &a Req
 &is ambigious?
 &ontology-based
 &rule based
 &Tokenizer, POS Tagger, Lemmatizer, parser, domain-specific  corpus based on wiki and extracted nouns
 &a set of rules and heuristics
 &20  Reqs  documents
 &P: 0.80; R: 0.89
    \\
    \hline
  Wang et al.\cite{wang2013automatic}
 &Ambiguity Detection
 &a set of Reqs
 &Overloaded and Synonymous ambiguity
 &lexical and syntactic features
 &rule based
 &lexical features, contextual features, pattern based features
 & C-value method to extract candidate concepts
 &4 projects (459 reqs)
 &Mean average precision : 0.52-0.57
    \\
    \hline
 Osama et al. \cite{osama2020score}
 &Ambiguity Detection
 &a Req
 &resolving syntactic ambiguity 
 &lexical and syntactic features
 &rule based
 &syntactic parsing
 & set of heuristics 
 &126 requirements
 &P:0.65 precision, R:0.99
    \\
    \hline
 Misra et al \cite{misra2013entity}
 &Ambiguity Detection
 &a Req
 &aliases disambiguity
 &Topic modeling
 &rule based
 &POS Tagging, terms extraction, stop words removal, misspelling identification, aliases identification, LSA
 &Similarity-based approach
 &65 reqs
 & F1: 0.6
    \\
    \hline
 Ferrari et al. \cite{ferrari2012using}
 &Ambiguity Detection
 &a set of Reqs
 &detect ambiguity
 &ontology-based
 &rule based
 &Build a domain related knowledge graphs
 &takes the concepts, searches for concept paths within each graph, compare paths
 &proof of concept with an example
 &-
    \\
    \hline
 Mishra et al. \cite{mishra2019use}
 &Ambiguity Detection
 &2 corpuses: related/not related to the domain
 &domain specific ambiguous CS words
 &advanced embedding
 &rule based
 &Tokenization, Punctuation  Removal,Stop word removal, PoS Tagging, Lemmatization, learn word2vec from each corpus
 &find out which of the commonly used CS words have a different meaning
 &a set of examples
 &-
    \\
    \hline
 matsuoka et al. \cite{matsuoka2011ambiguity}
 &Ambiguity Detection
 &SRS
 &detect ambiguous terms
 &ontology-based
 &rule based
 &sentence tokenization, extract terms using wordNet, C-Value
 &For each term: get related sentences, and use C-value and WordNet semantic similarity to cluster them 
 &a set of examples
 &-
    \\
    \hline
Ferrari et al. \cite{ferrari2019nlp}
 &Ambiguity Detection
 &corpuses from different domains
 &ambiguous terms
 &advanced embedding
 &rule based
 &crawl Wikipedia to extract domain specific documents; Language Models Generation
 &For each domain: find most freq. nouns, find a set of top similar words for each noun. Compare similar words to find ambiguity score
 &data from 5 domains
 &Maximum Kendall’s Tau of 88%
    \\
    \hline
    Dalpiaz et al. \cite{dalpiaz2019detecting}
    &Ambiguity Detection
 &a set of Reqs
 &near-synonyms terms
 &lexical and syntactic features
 &rule based
 &lexical and syntactic
 &calculate the similarity between terms then between their context using Cortical.io
 &28 data sets
 &R: 0.25; P: 0.51
    \\
    \hline
 Parra et al. \cite{parra2015methodology}
 &General Assessment
 &a set of reqs
 &Good or Bad Requ
 &lexical and syntactic features
 &ML
 &24 quality metrics that reflect lexical and syntactic information
 &C4.5*, PART, with Bagging and Boosting
 &INCOSE corpus
 &Acc: 0.87
    \\
    \hline
 Femmer et al. \cite{femmer2017rapid}
 &General Assessment
 &a set of Reqs
 & quality scores
 &lexical and syntactic features
 &rule based
 &POS Tagging; Lemmatization
 &a set of rules and heuristics
 &1471 reqs
 &P: 0.59; R:0.82
    \\
    \hline
 Ferrari et al. \cite{ferrari2018detecting}
 &General Assessment
 &a Requirement
 &10 types of defects
 &lexical and syntactic features
 &rule based
 &Tokenization; POS Tagging; Shallow Parsing; Gazetters; JAPE rules; 
 &dictionaties and rules
 &D-Pilot  and D-Large
 &P: 0.83; R:0.85
    \\
    \hline
 Ferrari et al. \cite{ferrari2014measuring}
 &Incompleteness Detection
 &a set of Reqs
 &Completeness score
 &lexical and syntactic features
 &rule based
 &POS Tagging, extract POS-based patterns (adjective+noun)
 &calculate completeness score based on C-NC Values
 &pilot study
 &-
    \\
    \hline
 Liu et al. \cite{liu2021automated}
 &Incompleteness Detection
 &SRS Document
 &is incomplete?
 &advanced embedding
 &ML
 &Tokenization, POS, Stop words removal; Lemmatization; skip-thoughts sent2vec encoder
 &a graph-based clustering
 &a set of aerospace reqs
 &F1:0.52
    \\
    \hline
 Baumer et al. \cite{baumer2018flexible}
 &Incompleteness and Ambiguity Detection
 &a Req
 &solve predefined defect cases
 &ontology-based
 &rule based
 &Semantic Labeling; Stopwords removal; POS tagging; WordNet
 & a set of predefined rules to solve some defect cases 
 &400 selected requirements
 &F1: Incomp.(0.76); Ambig. (0.78)
    \\
    \hline
 Cruz et al. \cite{cruz2017detecting}
 &Vagueness Detection
 &a Req
 &is vague?
 &lexical and syntactic features
 &rule based
 &sentence splitrer, POS-tagging, get adjectives and adverbs that match a blacklist of known vague terms or does not match a whitelist of not vague terms.
 &pilot study
 &P: 0.34; R:0.94
    \\
    \hline
 Arora et al. \cite{arora2015automated}
 &conformance with templates
 &a Req
 &Conformance or Not
 &lexical and syntactic features
 &rule based
 &Tokenization; POS tagging; NER; Chunking
 &Req Exp based on pattern matching 
 &4 test cases
 &F2: 0.96-1.0
    \\
    \hline
 Abualhaija et al. \cite{abualhaija2020automated}
 &Extraction from Docs
 &a textual document
 &Req or Not Req
 &lexical and syntactic features
 &ML
 &20 Features: token-based features, syntactic features, semantic features, frequency-based features
 &RF with cost sensitive learning + a set of Post-Processing steps
 &33 industrial set of reqs.
 &P: 0.81, R: 0.95
    \\
    \hline
Haris et al \cite{haris2020automated}
 &Extraction from Docs
 &SRS
 &Req or Not Req
 &lexical and syntactic features
 &rule based
 &POS Tagging, chunking
 &POS tagging pattern
 &PURE dataset
 &P:0.64-1.0; R: 0.64-0.89
    \\
    \hline
 Quirchmayr et al. \cite{quirchmayr2018semi}
 &Extraction from Docs
 &user manual
 &list of features
 &lexical and syntactic features
 &rule based
 &POS tagging; terms extraction ; parsing; Pattern-based parse tree transformations and correction
 &Syntactical patterns
 &industrial case
 &F1:0.92
    \\
    \hline
  Wang et al. \cite{wang2013automatic}
 &Extraction from Docs
 &a textual document
 &requirements
 &lexical and syntactic features
 &ML
 &Tokenization, lemmatization, pos tagging, dependency parsing
 &Bi-LSTM-CRF
 &22 SRS documents
 &F1: 0.86
    \\
    \hline
 Shi et al. \cite{shi2021automatically}
 &Extraction from Docs
 &Mailing list
 &Feature requests
 &lexical and syntactic features
 &rule based
 &81 fuzzy rules
 &10 semantic sequence patterns
 &317
emails from Ubuntu community 
 &Avg precision: 0.76
Avg recall: 0.86
    \\
    \hline
 De et al. \cite{de2021re}
 &Extraction from Reviews
 &user review
 &extract chunk which represent a feature
 &advanced embedding
 &ML
 &tokenization, BERT
 &use B,I,O annotations to label boundaries
 & reviews dataset for 8 Apps (125 for each)
 &F1 Exact Match (0.46); 
Partial Match (0.62)
    \\
    \hline
  Bakar et al. \cite{bakar2016extracting}
 &Extraction from Reviews
 &user review
 &features
 &Topic modeling
 &rule based
 &stop words removal, lemmatization, pos tagging, TF-IDF, Term-Document Matrix, 
 & syntactic patterns to extract features
 & 32 software / 1253 sentences
 &P: 0.87 (0.62 avg); R: 0.86 (0.82 avg)
    \\
    \hline
 Carreno et al. \cite{carreno2013analysis}
 &Extraction from Reviews
 &user feedback
 &main topics
 &Topic modeling
 &ML
 &tokenization, stop words removal, 
 &topic modeling and sentiment analysis
 &data set of 327 comments
 &P: 0.90; R: 0.5-0.8
    \\
    \hline
 Jha et al. \cite{jha2018using}
 &Extraction from Reviews
 &user review
 &bugs or features
 &VSM
 &ML
 &(1) BOF: based a probabilistic frame semantic parser. (2) BOW: Stemmer, BOW
 &SVM*, NB
 &Three datasets of reviews
 &F1: Features (0.75); Bugs (0.85)
    \\
    \hline
  Li et al. \cite{li2018automatically}
 &Extraction from Reviews
 &user requests
 &classify user requests to 8 types
 &lexical and syntactic features
 &ML
 &Unigram; TF-IDE; lexical and syntactic patterns
 &SVM*, KNN, NB
 &KeePass, Mumble, WinMerge
 &Avg Acc: 0.77; Avg R: 0.54-0.81; Avg P:0.73-0.92
    \\
    \hline
    Lu et al. \cite{lu2017automatic}
 &Extraction from Reviews
 &user review
 &4 reviews classes.
 &advanced embedding
 &ML
 &the paper used word2vec to produce: AUR-BoW
 &Naive Bayes, J48, and Bagging*
 &4000 sentences 
 &F1: 0.71
    \\
    \hline
 Peng et al. \cite{peng2016approach}
 &Extraction from Reviews
 &user reviews
 &feature requests
 &lexical and syntactic features
 &ML
 &lexical processing, pos tagging, dependency parser
 &Naive Bayes
 &1924 reviews 
 &F: 0.82
    \\
    \hline
 Yang et al. \cite{yang2021phrase}
 &Extraction from Reviews
 &user reviews
 &features
 &lexical and syntactic features
 &ML
 &Sentence splitting, sentiment analysis,
POS tagging, stemming
 &RF
 &reviews for 6 products
 &F1: 0.72
    \\
    \hline
  Wu et al. \cite{wu2021identifying}
 &Extraction from Reviews
 &reviews
 &key features
 &advanced embedding
 &ML
 &tokenization, pos tagging, dependency parsing, BERT 
 & A regression  model  to  identify  features
 & 200 app with 1,1M reviews.
 &F1: 0.78
    \\
    \hline
  Jiang et al. \cite{jiang2019recommending}
 &Extraction From Similar Apps
 &featurs+repo. of similar applications
 &suggested new features
 &Topic modeling
 &ML
 &BOW+LDA
 &clustering technique
 &533 annotated features from 100 apps
 &Hit@15 score: up to 78\%
    \\
    \hline
 Abbas et al.\cite{abbas2020automated}
 &Extraction From Similar Apps
 &corpus of reqs
 &recommended requirement
 &VSM
 &ML
 &Stop words removal, pos tagger, Lemmatization, TF-IDF
 &clustering requirements+ use Neighbours to recommend reqs
 &188 requirements
 &Acc: 0.74
    \\
    \hline
 Do et al. \cite{do2020capturing}
 &Extraction From Similar Apps
 &set of requirements
 &suggested new requirements
 &advanced embedding
 &ML
 &stemming, tokenization, doc2vec, POS tagging
 &clustering based BIRCH algorithm
 &571 systems from 3 domains
 &Human evaluation
    \\
    \hline
  Hussain et al. \cite{hussain2013approximation}
 &Effort Estimation
 &a set of reqs
 &Function Point
 Level
 &lexical and syntactic features
 &ML
 &features including: 
NP; Parentheses; Active Verbs; Tokens in parentheses; Conjunctions; Pronouns; VP;  Words; Sentences; Uniques (hapax legomena)
 &C4.5
 &4 projects
 &F1: 0.66
    \\
    \hline
  Choetkiertikul et al.\cite{choetkiertikul2018deep}
 &Effort Estimation
 &requirements
 &story points
 &advanced embedding
 &ML
 &word embedding
 &LSTM and recurrent highway network
 &23,313 issues
 &outperforms 3 common baselines
    \\
    \hline
  Ali et al. \cite{ali2021requirement}
 &Prioritization
 &set of requirements
 &priority
 &VSM
 &rule based
 &Tokenization, POS tagging, Stemming, Stop words removal,BOW
 &case-based reasoning (CBR) technique
 &four different case studies
 &
    \\
    \hline
  Kifetew et al. \cite{kifetew2021automating}
 &Prioritization
 &set of reqs + reqs' feedbacks
 &priority
 &lexical and syntactic features
 &rule based
 &tokenization, pos tagging, stemming, sentiment analysis, intention score, severity score
 &extracts quantifiable properties relevant for prioritizing reqs.
 &two empirical studies
 &
    \\
    \hline
  McZara et al. \cite{mczara2015software}
 &Prioritization
 &set of requirements
 &priority
 &lexical and syntactic features
 &rule based
 &tokenization, stemming, pos tagging
 &SMT solver + AHP
 &100 reqs (top 20 are annotated)
 &
    \\
    \hline
Misra et al. \cite{misra2014latent}
 &Prioritization
 &set of requirements
 &priority
 &Topic modeling
 &rule based
 &pos tagging, entities and action extraction, LSA
 &similarity-based approach
 &41 requirements
 &
    \\
    \hline
  Shafiq et al. \cite{shafiq2021nlp4ip}
 &Prioritization
 &a set of Reqs
 &suggested priority
 &lexical and syntactic features
 &ML
 &
 &semi-automatic approach
 &19 projects 
 &average top\@3 acc 81\% 
    \\
    \hline
 Al-Hroob et al. \cite{al2018use}
 &Modeling (UML)
 &a set of Reqs
 &Use Cases
 &lexical and syntactic features
 &ML
 &Tokenization, pos tagging, dependency parsing
 &ANN to predict the role of each word
 &5 test cases
 &F1: 0.43-0.44
   \\
    \hline
 Elallaoui et al. \cite{elallaoui2018automatic}
 &Modeling (UML)
 &a set of Reqs
 &use cases
 &lexical and syntactic features
 &rule based
 &pos tagging
 &rule based
 &90 user stories
 &P: 0.87; R:0.85
  \\
    \hline
  Hamza et al \cite{hamza2019generating}
 &Modeling (UML)
 &a set of Reqs
 &use cases
 &lexical and syntactic features
 &rule based
 &Tokenization; POS tagging; Chunking; Grammar
patterns tagging
 &rule based
 &4 test cases
 &P: 0.72; R: 0.69
 \\
    \hline
  Tiwari et al. \cite{tiwari2019approach}
 &Modeling (UML)
 &a set of Reqs
 &use cases
 &lexical and syntactic features
 &rule based
 &POS Tagger, Dependency Tree
 &rule based
 &10 case studies.
 &Human Evaluation
\\
    \hline
  Haj et al. \cite{haj2021semantic}
 &Modeling (SBVR) 
 &a set of Reqs
 &SBVR 
 &lexical and syntactic features
 &rule based
 &Lemmatization, pos tagging, dependency tree, NER, 
 &rule based
 &3 case studies
 &F1: 0.87
\\
    \hline
  Lucassen et al. \cite{lucassen2017extracting}
 &Modeling (Conceptual Model)
 &a set of Reqs
 &concepts and relations
 &lexical and syntactic features
 &rule based
 &tokenization; pos tagging, dependency parsing
 &a set of heuristic rules
 &4 industrial data sets
 &R:0.88-0.97; P:0.92-0.98
\\
    \hline
  Thakur et al. \cite{thakur2016identifying}
 &Modeling (Conceptual Model)
 &a set of Reqs
 &concepts and relations
 &lexical and syntactic features
 &rule based
 &tokenization; pos tagging, dependency parsing
 &rule based
 &40 use cases
 &Acc: 0.91
\\
    \hline
  Hamza et al \cite{hamza2015recommending}
 &Modeling (Features Model)
 &a set of Reqs
 &Features Model
 &lexical and syntactic features
 &rule based
 &tokenization, pos tagging
 &a set of heuristic rules
 &4 case studies
 &human evaluation
\\
    \hline
  Sree-Kumar et al. \cite{sree2018extracting}
 &Modeling (Features Model)
 &a set of Reqs
 &Features
 &lexical and syntactic features
 &rule based
 &Stop words removal, Tokenization, NP chunking, TF-IDF
 &Set of heuristics rules
 &6 case studies
 &Features: P 0.40-0.73 / R 0.57-0.93
Relation: P 0.41-0.87 / R 0.48-0.76
\\
    \hline
  Arora et al.\cite{arora2016automated}
 &Glossary Terms Extraction
 &a set of Reqs
 &Glossary + clustering terms
 &ontology-based
 &rule based
 &tokenization, pos tagging, chunking, Soft TF-IDF, wordNet
 & a set of huristics + similarity-based approach
 &3 sets each consists of 110-380 reqs
 &F: 0.52 - 0.64
\\
    \hline
  Bhatia et al. \cite{bhatia2020clustering}
 &Glossary Terms Extraction
 &a set of Reqs
 &Glossary + clustering terms
 &advanced embedding
 &ML
 &Stop words removal, Tokenization, chunking; Lemmatization, WordNet to remove common concepts
 &Similarity Matrix is calculated based on fastText + EM for clustering
 &CrowdRE Dataset
 &F1: Terms (0.39); Clustering (0.65)
\\
    \hline
  Gunes et al \cite{gunecs2020automated}
 &Modeling (Goal Model)
 &a set of Reqs
 &goal model
 &ontology-based
 &rule based
 &stopwords removal, pos tagging, extract user story elements,  similarity scores
 &huristic rules
 &a test case
 &-
\\
    \hline
  Mishra et al. \cite{mishra2021crawling}
 &Language Model
 &corpus of texts from wiki
 &SE word embeddings model
 &advanced embedding
 &ML
 &word embedding
 &word2vec
 &size of our trained model is 92MB
 &-
 \\
    \hline
  Dollmann et al. \cite{dollmann2016and}
 & Extraction From Docs + Semantic Labeling
 &a set of sentences
 & reqs+Semantic labeling
 &lexical and syntactic features
 &ML
 &Features: (1) Extraction Approach (BOW; Length; dependencies; POS tags).
Semantic Labeling (Orthographic; Semantic using  WordNet; dependencies)
 &ExtraTreeClassifier
 & dataset from SourceForge
 &F1: Extraction(0.91); semantic labeling (0.73)
\\
    \hline
  Sainani et al. \cite{sainani2020extracting}
 &Extraction From Docs / Classification
 &Contracts documents
 &Obligation or not
 &advanced embedding
 &ML
 &stop words, lemmatization, bigrams and trigrams, TF-IDF
 &BiLSTM
 &contracts document (250 pages, 1608 sentences)
 &F1: Extraction (0.93); Classification (0.60-0.96)
\\
    \hline
        \caption{The Detailed Info for all 104 selected papers}
        \label{tab:t3}
        \end{longtable}
\end{landscape}

\bibliography{elsarticle-template}

\end{document}